\documentclass[12pt]{article}
\usepackage{jheppub}

\def\CN {{\cal N}}
\def\CR {{\cal R}}

\def\CF {{\cal F}}

\def\CP {{\cal P }}

\def\CV {{\cal V}}
\def\CW {{\cal W}}

\def\CO {{\cal O}}
\def\CZ {{\cal Z}}

\def\CB {{\cal B}}
\def\CS {{\cal S}}

\def\CZ{{\cal Z}}
\newcommand{\Tr}{{\rm Tr\,}}
\def\IC {{\mathbb C}}

\title{Surface Defects and Resolvents}

\author[1,2]{Davide Gaiotto}
\author[3]{Sergei Gukov}
\author[2]{Nathan Seiberg}
\affiliation[1]{Perimeter Institute for Theoretical Physics, Waterloo, Ontario, Canada N2L 2Y5}
\affiliation[2]{Institute for Advanced Study, Einstein Dr., Princeton, NJ 08540, USA}
\affiliation[3]{California Institute of Technology, Pasadena, CA 91125, USA}

\abstract{We study a large class of BPS surface defects in $4d$ $\CN =2$ gauge theories.  They are defined by coupling a $2d$ $\CN=(2,2)$ gauged linear sigma model to the $4d$ bulk degrees of freedom.  Our main result is an efficient computation of the effective twisted superpotential for all these models in terms of a basic object closely related to the resolvent of the $4d$ gauge theory, which encodes the curve describing the $4d$ low energy dynamics. We reproduce and extend the results of brane constructions and compute the effective twisted superpotential for general monodromy surface defects. We encounter novel, puzzling field theory phenomena in the low energy dynamics of the simplest surface defects and we propose some local models to explain them. We also study in some detail the behavior of surface defects near monopole points of the bulk theory's Coulomb branch.  Finally, we explore the effect on the defect of breaking the bulk supersymmetry from $\CN=2$ to $\CN=1$ and show that certain quantities are independent of this breaking.
}

\begin{document}

\maketitle

\section{Introduction}

BPS surface defects (a.k.a. surface operators) in $\CN=2$ four-dimensional gauge theories are an interesting theoretical subject
which ties together many subjects in mathematical physics \cite{Gukov:2006jk,Gomis:2007fi,Gukov:2007ck,Alday:2009fs,Gaiotto:2009fs,
Kapustin:2010pk,Kozcaz:2010af,Alday:2010vg,Dimofte:2010tz,Awata:2010bz,
Kozcaz:2010yp,Wyllard:2010rp,Gaiotto:2011tf,Kanno:2011fw,Nakayama:2011pa,
Longhi:2012mj,Gaiotto:2012xa,Iqbal:2012xm,Alday:2013kda,Babaro:2013yj,
Dorey:2011pa,Bonelli:2011fq,Wyllard:2010vi,Yamada:2010rr,Taki:2010bj,Tan:2009qq,Tan:2010dk}. The interplay between four-dimensional and two-dimensional degrees of freedom allows for intricate dynamics, which we can probe through exact calculations.

A given four-dimensional $\CN=2$ gauge theory can be modified in the UV in a  large variety of ways to define distinct half-BPS surface defects, which preserve two-dimensional
$(2,2)$ supersymmetry  \cite{Gukov:2006jk}. The simplest example of half-BPS surface defect is a GW surface defect, which is defined in a way, which is akin to the definition of a 't Hooft loop: the UV path integral is done on gauge field configurations with a non-zero semi-simple monodromy on a small loop around the defect. The choice of
monodromy breaks the gauge symmetry at the defect to some (``Levi'') subgroup $L$ of the gauge group, which labels the type of defect.
Another general strategy is to couple a two-dimensional theory with $(2,2)$ supersymmetry to four-dimensional gauge fields, by gauging a $2d$ flavor symmetry.\footnote{Although a similar-looking 2d-4d coupled system describes degrees of freedom of half-BPS vortex strings
(see {\it e.g.} \cite{Dorey:1998yh,Hanany:2003hp,Auzzi:2003fs,Shifman:2004dr,Hanany:2004ea,Shifman:2008wv} for a small sample of work on this subject),
it is important to understand a distinction between dynamical vortices and surface defects.  This distinction is analogous to the distinction between dynamical charged particles and the world lines of probe particles. The latter define line operators, such as Wilson or 't Hooft lines, and can be introduced in a gauge theory even if there are no dynamical particles with the same charges. Similarly, in a given gauge theory, the surface defects represent the world-sheets of probe strings.  Their spectrum is often larger than the possible dynamical vortices. A simple example that will be extensively discussed in this paper is a pure super-Yang-Mills theory, which has no half-BPS vortices at all, but admits a large spectrum of surface defects.}
It is possible to mix the two strategies, coupling $2d$ degrees of freedom to a factor in $L$.

As we flow to the IR, the bulk degrees of freedom reduce generically to an Abelian theory, fully characterized by an effective pre-potential $\CF$,
which controls the geometry of the moduli space $\CB$ of Coulomb branch vacua \cite{Seiberg:1994rs}. The surface defect will generically have a set of
massive vacua, fibered in a non-trivial way over the Coulomb branch of the four-dimensional theory to form a space of vacua $\CV$ for the combined $2d$-$4d$ system.
In each massive vacuum, the surface defect should reduce to a simple monodromy defect in the Abelian gauge theory, which is fully characterized by an effective twisted superpotential $\CW$ \cite{Alday:2009fs}.
The choice of surface defect in the UV determines the geometry of the space of vacua, and the effective superpotential in the IR.

There is an intricate connection between the analytic properties of the IR effective twisted superpotential for a surface defect
and the pre-potential of the bulk theory \cite{Alday:2009fs,Gaiotto:2009fs,Gaiotto:2011tf}. The pre-potential itself, of course, has a complicated analytic structure: if we follow a generic closed path in the Coulomb branch, the low energy Abelian theory may come back to itself up to an electric-magnetic
duality transformation, which acts appropriately on the pre-potential. In particular, it acts as a constant symplectic transformation on the periods
$(a, a_D)$ of the bulk theory \cite{Seiberg:1994rs}.

The effective twisted superpotential $\CW$ of a two-dimensional $(2,2)$ theory, on the other hand,
has a relatively simple analytic structure over the space of vacua of the $2d$ theory: it can only shift
by an integer linear combination of the twisted mass parameters, which are in one-to-one correspondence with the flavor symmetries of the $2d$ theory \cite{Hanany:1997vm,Dorey:1998yh}.
In order to understand the analytic properties of the effective twisted superpotential $\CW$ for a surface defect \cite{Alday:2009fs,Gaiotto:2009fs},
we can imagine constructing the defect by weakly coupling the flavor symmetry of a two-dimensional $(2,2)$ theory
to four-dimensional gauge fields. Semiclassically, the twisted mass parameters are replaced by the electric periods $a$.
As we explore the Coulomb branch, the periods mix, and $\CW$ in a given vacuum is defined up to shifts by a linear combination with integer coefficients of
$a$ and $a_D$.

This statement agrees with the expected IR description of surface defects. The electric $\alpha$ and magnetic $\eta$
monodromy parameters \cite{Gukov:2006jk} are computed from the gradient of $\CW$ along the Coulomb branch as
\begin{equation} \label{eq:dCW}
d \CW = \eta \cdot da + \alpha \cdot d a_D ~.
\end{equation}
Shifts of $\CW$ by $n_e  \cdot a + n_b  \cdot a_D$ correspond to the natural periodicities of the monodromy parameters.
This beautiful picture is, in a sense, deeply puzzling \cite{Gaiotto:2009fs}. We can start from some completely generic massive $(2,2)$
theory, with a flavor symmetry $G$ and an arbitrarily complicated low energy effective twisted superpotential,
and couple it to some bulk theory with gauge group $G$. Magically, the coupling must correct the geometry of the space of vacua
and the effective twisted superpotential just in the right way, so that $\CW$ acquires discontinuities proportional to the
periods of the bulk theory and evolves along the Coulomb branch accordingly to the correct geometry.

Brane constructions \cite{Hanany:1997vm} or six-dimensional engineering \cite{Alday:2009fs,Gaiotto:2011tf}
often provides us with a handful of natural surface defects
and exact expressions for the low-energy quantities which match these expectations,
but do not shed light on the precise field-theory mechanism behind it.
The purpose of this paper is to reproduce these results directly from a field theory analysis of the problem
and extend them to more general systems that do not admit a brane engineering construction.
We aim to clarify several puzzling features of the surface defects low energy dynamics and to
identify which low energy degrees of freedom are relevant in various corners of their parameter space.

Our main computational tool is an extension of the powerful two-dimensional gauged linear sigma model technology \cite{Witten:1993yc,Hanany:1997vm},
which in principle allows one to compute the effective twisted superpotential for all surface defects
defined by coupling the bulk gauge fields to flavor symmetries of two-dimensional GLSMs.
This large class of defects includes in particular both all brane engineering examples and
the general GW surface defects.

At first we focus on surface defects with a full UV definition. We take a UV description of the bulk theory,
with gauge group $G$, and couple it to a two-dimensional $(2,2)$ GLSM with a global symmetry $G$ (If $G$ is a product group, part of it can act trivially).
The simplest choice of $2d$ theory is a set of $2d$ chiral matter fields
in some representation of the bulk gauge group $G$.
For example, if our bulk gauge group is $G=SU(N)$, the $2d$ surface defect theory can be a theory of $N$ free chiral superfields.  The bulk scalars appear as twisted mass terms for the $N$ fields, and an additional complex parameter $m$
is a common $2d$ twisted mass for all these fields.
We learn how to integrate out the $2d$ chiral fields to derive an effective twisted superpotential
${\cal W}(u, m)$, a locally holomorphic function of the 4d Coulomb branch parameters $u$ and $2d$ twisted masses $m$.

We can make this surface defect more complicated by adding $2d$ gauge fields coupled to some other global symmetries of the $2d$ chiral fields. In our example this will be the global $U(1)$ symmetry associated with the parameter $m$.
Now $m$ is the expectation value of a dynamical field $\sigma$.  We can add an FI term $t$ for the $U(1)$ gauge field
and thus engineer a $\mathbb{CP}^{N-1}$ GSLM coupled to the bulk $SU(N)$ gauge theory. The effective twisted superpotential ${\cal W}(u, t)$ is simply the Legendre transform of the previous ${\cal W}(u, m)$
for the fundamental representation of $SU(N)$.

Finally, we can add extra $2d$ chiral fields and $2d$ gauge fields to the system to engineer a general GLSM coupled to the bulk gauge fields. We can also add couplings to the bulk hypermultiplets. As long as we know the original ${\cal W}(u, m)$, the calculation of the effective twisted superpotential for the resulting theory is still elementary.

Alternatively, we can take a macroscopic view of the problem.  In the bulk the gauge group $G$ is replaced at low energies by an Abelian gauge group $H\subset G$.  The low energy theory is an $H$ gauge theory and is typically non-renormalizable -- it includes higher dimension operators.  We can adopt a similar point of view about the defect.  Here we need a two-dimensional system with a global symmetry $H$.  And as in the bulk, in order to match to a particular UV theory we must add to this simple $2d$ Lagrangian various high dimensional operators. For example, if $G=SU(2)$, then $H=U(1)$, and we can consider a surface defect with a single chiral superfield $Z$.  This simple model has only the symmetry $H$ and cannot include additional gauge fields.  In order to make it agree with the more complete microscopic surface defect, we need to add to its simple Lagrangian certain high dimension operators.  As we will see, this is most easily done after $Z$ is dualized.  A more complicated model can be constructed by having several chiral superfields with various $U(1)$ charges.  Then, we can also add gauge fields to the model.

We now discuss the structure of the paper. In section 2 we show how to compute the effective twisted superpotential for surface defects defined by gauged linear sigma models. In section 3 we demonstrate these ideas by studying several examples.  We consider different defects coupled to a pure $SU(2)$ gauge theory and then to more complicated theories with larger gauge groups and with matter fields in various representations.  In section 4 we consider effective local models valid only in some regions of the moduli space of vacua.  This region can include singularities like the monopole points on the Coulomb branch.  In section 5 we prepare the ground to breaking $\CN=2$ to $\CN=1$ supersymmetry by identifying the appropriate operator describing this breaking.  Section 6 addresses the effect of this breaking of supersymmetry on the defect.  Surprisingly, we find certain non-renormalization theorems (other non-renormalization theorems were discussed in \cite{Cui:2011rz}).  Here we identify some of the interesting observables in the field theory in terms of the brane constructions of these models.  Finally, in section 7 we summarize our results and offer an outlook.

\section{Surface defects at generic points in the Coulomb branch}
In this section we propose a description of the IR physics in a generic massive vacuum for
a large class of surface defects in $\CN=2$ gauge theories,
which are defined by coupling a two-dimensional $(2,2)$ gauged linear sigma model
with a flavor symmetry $G$ to $G$-valued four-dimensional gauge fields.

It is useful to remember the general procedure to compute the low energy twisted effective superpotential for a gauged linear sigma model \cite{Witten:1993yc,Hanany:1997vm}.
It involves two steps:\begin{itemize}
\item Integrate out the $2d$ chiral matter fields. A chiral field of total mass $x$, including contributions from twisted masses $m$ and the $2d$ gauge multiplet scalars $\sigma$, gives an effective twisted superpotential $2 \pi i \CW[x] = -x \log x/e$.\footnote{In this formula and in the rest of the paper we omit an obvious $2d$ renormalization scale. It can be reinstated easily by dimensional reasoning and/or absorbed in the renormalized FI parameters. Appropriate powers of the $2d$ renormalization scale would appear in the arguments of most logarithms throughout the paper.}
\item Extremize the total effective superpotential $\CW[\sigma, m]$ with respect to the gauge multiplet scalar vevs. The extremum equations give twisted chiral ring relations for the theory.
The extremum values correspond to the low energy twisted effective superpotential in the various massive vacua of the theory.
\end{itemize}

The canonical example is the $\mathbb{CP}^1$ two-dimensional gauged linear sigma model, defined by a $U(1)$ gauge theory with two chiral fields of charge $1$,
which transform as a doublet for an $SU(2)$ flavor symmetry.
After integrating out the chirals, we get
\begin{equation} \label{eq:cp1}
2 \pi i\CW[\sigma, m] = t \sigma - (\sigma - m) \log(\sigma - m)/e - (\sigma + m) \log (\sigma + m)/e
~, \end{equation}
where $m$ is a twisted mass parameter associated with the flavor $SU(2)$ symmetry.
The extremum equations become the standard twisted chiral ring relation:
\begin{equation}
\sigma^2 - m^2 = e^t
~. \end{equation}
There are two massive vacua, and the low energy effective superpotential in each vacuum can be derived
by substituting the solutions of the twisted chiral ring relation into $\CW[\sigma, m]$.

We propose a similar procedure for surface defects:
\begin{itemize}
\item Integrate out the chiral matter fields. Treat the bulk vectormultiplet scalars $\Phi$ as twisted masses at this stage.
Thus a $2d$ chiral field of $2d$ mass $x$, in a representation $\CR$ of the bulk gauge fields, gives an effective twisted superpotential $2\pi i\CW_\CR[x,\Phi] = -\mathrm{Tr}_{\CR} (x+\Phi) \log (x+\Phi)/e$.
\item Replace each $\mathrm{Tr}_{\CR} \Phi^k$ in the expansion of $\CW_\CR[x,\Phi]$ with an appropriate polynomial $p^{(k)}_{\CR}[u_i]$ of the Coulomb branch parameters of the bulk theory $u_i$. The result is an intermediate twisted superpotential $\CW[\sigma, m,u]$, which is a function of $2d$ scalar vevs $\sigma$, $4d$ scalar vevs $u_i$, and mass parameters $m$.
\item Extremize the total effective superpotential $\CW[\sigma, m,u]$ with respect to the $2d$ gauge multiplet scalar vevs. The extremum equations give twisted chiral ring relations for the surface defect.
\end{itemize}

In order to make our proposal concrete, we need to be specific about the map $\mathrm{Tr}_{\CR} \Phi^k \to p^{(k)}_{\CR}[u_i]$.
If $\CR$ is the fundamental representation for a unitary gauge group,
there are three possible independent ways to determine the result, which appear to give the same answer: \begin{itemize}
\item We can pick a brane system that engineers simple gauged linear sigma model surface defects, and solve them through an M-theory lift.  The solution gives the exact chiral ring relations, from which we can reverse-engineer the desired map.
\item We can identify the relations $\mathrm{Tr}\Phi^k = p^{(k)}[u_i]$ as the quantum chiral ring relations for the bulk theory, which were derived in \cite{Cachazo:2002ry,Seiberg:2002jq,Cachazo:2003yc} (following the work of \cite{Dijkgraaf:2002fc,Dijkgraaf:2002vw,Dijkgraaf:2002dh}) through a careful analysis of the generalized Konishi anomaly equations in field theory.
\item We can use localization computations, as in \cite{Nekrasov:2003rj}: take the $\epsilon_i \to 0$ limit of the vev of $\mathrm{Tr}\Phi^k$ in an $\Omega$ background.
\end{itemize}

It would take some extra work to figure out the answer for general representations of an unitary gauge group.
The simplest option would probably be to generalize the known localization results.\footnote{Notice that although we can rewrite $\mathrm{Tr}_{\CR} \Phi^k$ as a polynomial in elementary fundamental traces,
we cannot simply replace the traces in that expression by the $p^{(k)}[u_i]$. In the $\Omega$ background,
Coulomb branch operators preserve SUSY only when inserted at the origin, and thus the vev of a product
of operators is not the same as the product of the vevs.}
For other gauge groups, localization is more cumbersome. The M-theory trick should provide some useful answers for the fundamental
representations of orthogonal or symplectic groups.

\section{Basic examples}
\subsection{Pure $SU(2)$}
We will start with a few examples of surface defects for a pure $SU(2)$ gauge theory.
The best known example is the GW monodromy defect, defined by breaking $SU(2)$ to $U(1)$ at the defect.
This defect has a dual UV definition as a $\mathbb{CP}^1$ $2d$ sigma model coupled to the pure $SU(2)$ theory \cite{Gukov:2006jk,Kanno:2011fw}. As the $\mathbb{CP}^1$ $2d$ sigma model has a GLSM description, we define the defect in terms of
an $SU(2)$ doublet of $2d$  chiral fields,
of charge $1$ under a $2d$  $U(1)$ gauge field and use our tools to analyze it.
It should be clear that from our point of view, this is not the simplest possible example: it is more natural to only include the $2d$ chiral doublet,
with an ungauged $2d$ $U(1)$ flavor symmetry. This is a very interesting defect, which can be used as a universal building block to construct more complicated systems.
As the properties of such basic defect are somewhat counter-intuitive, we will first describe the more familiar $\mathbb{CP}^1$ example, and then present this more elementary model.

\subsubsection{Coupling to $\mathbb{CP}^1$}
If we turn off the four-dimensional gauge dynamics, the four-dimensional vectormultiplet scalar field $\Phi$
becomes a twisted mass for the $SU(2)$ flavor symmetry of the two-dimensional model. In particular, the eigenvalues of $\Phi$ play the same role as
the $SU(2)$ twisted mass parameter $m$ in (\ref{eq:cp1}). We can write the
twisted effective superpotential $\CW(\sigma,\Phi)$ in a matrix form as
\begin{equation}\label{eq:cwP}
2 \pi i\CW[\sigma,\Phi] =t \sigma - \mathrm{Tr} \left( \sigma + \Phi \right) \log \left( \sigma + \Phi \right)/e
~. \end{equation}

How do we introduce the effects of the four-dimensional gauge dynamics?
If we take two $\sigma$ derivatives of $\CW[\sigma,\Phi]$, we find a familiar object
\begin{equation}
-2 \pi i \partial_\sigma^2 \CW[\sigma,\Phi] = \mathrm{Tr} \frac{1}{\sigma + \Phi} ~.
\label{Wreslll}
\end{equation}
The object on the right hand side plays a crucial role both in localization calculations, where it is usually denoted as $R(\sigma)$,
and in calculations involving $\CN=1$ anomaly equations, where it is usually denoted as $T(\sigma)$.
It is the generating function for the traces $\mathrm{Tr} \Phi^k$.
We will use the $T(\sigma)$ terminology here, to avoid confusion with another object $R(\sigma)$, which appears in
the $\CN=1$ context. The origin of the conflicting notation is that one $R(\sigma)$ is the resolvent of the Nekrasov-Okounkov matrix integral,
while the other is the resolvent of the Dijkgraaf-Vafa matrix integral.

The expression for $T(x)$ from either computational setup is
\begin{equation}
T(x) \equiv  \mathrm{Tr}\frac{1}{x + \Phi}  = \frac{2 x}{\sqrt{(x^2- u)^2 - 4 \Lambda^4}} =\frac{2}{x} + \frac{2 u}{x^3} + \frac{2u^2 + 4 \Lambda^4}{x^5}+ \cdots
~. \end{equation}
We see the systematic instanton corrections to the naive $\mathrm{Tr} \Phi^{2k} = 2 u^k$.

In order to derive the twisted chiral ring equation, we need to integrate $T(\sigma)$ once, to
\begin{equation}
2 \pi i \partial_\sigma \CW[\sigma,u] =t - \mathrm{arccosh} \frac{\sigma^2- u}{2\Lambda^2} - \log \Lambda^2 = t - \log\left(\frac{\sigma^2- u + \sqrt{\left(\sigma^2- u\right)^2- 4\Lambda^4}}{2} \right)
~. \end{equation}
Thus we recover the same twisted chiral ring equation, which arises in the brane solution to the problem \cite{Alday:2009fs,Gaiotto:2009fs}:
\begin{equation}
\sigma^2 = e^{t} + u + \Lambda^4 e^{-t}
~. \end{equation}

At this point we can simply integrate $T(\sigma)$ twice to write down $\CW[\sigma,u]$ and insert the solution of the bulk chiral ring relation
to compute the exact low-energy effective twisted superpotential. This is a slight improvement on the brane solution of the problem,
which typically provides us with the twisted chiral ring equation, and tells us to compute the low energy effective twisted superpotential
by integrating the canonical one-form
\begin{equation}
\lambda = \sigma dt~.
\end{equation}
This gives a natural way to compute the difference between the twisted super potentials of two vacua of the theory, which controls the mass of BPS solitons in the theory, but does not immediately fix the complete $u$ dependence of the answer.

With the full solution at hand, we can propose the complete answer in a given vacuum
\begin{equation}
2 \pi i \CW = 2 e^{t/2} - \int_t^\infty \left[ \sigma(t') -e^{t'/2} \right]dt'~,
\end{equation}
where the contour ends at $\infty$ on the branch where $\sigma(t') \sim e^{t'/2}$.
This reproduces the expected large $\sigma$ behavior of $T(\sigma)$.
Our result essentially agrees with the careful guess made in
\cite{Gaiotto:2011tf}.

The contour integrals of $\lambda$ also allow to compute the discontinuities of the twisted superpotential along
paths that start and end in the same vacuum.  From general considerations \cite{Alday:2009fs,Gaiotto:2009fs} these
are known to be integer linear combinations of the periods of the bulk theory (and possibly the $2d$ twisted masses).
In particular, the twisted chiral ring equations and canonical differential for a generic defect play the same role as the curve and differential for the bulk theory.
The specific defect at hand has the neat feature that its matter content is such as to reproduce precisely the standard curve, in the same way as the curve is
derived from the resolvent in \cite{Nekrasov:2003rj,Cachazo:2002ry}.

It is worth spending a few more words on the relation between our considerations and the $\CN=1$ analysis of \cite{Cachazo:2002ry}.
Classically, the contour integrals
\begin{equation}
\oint d\sigma \sigma T(\sigma)  = \oint d\sigma \mathrm{Tr} \frac{\sigma}{\sigma + \Phi}
\end{equation}
pick the poles of the integrand, and give the eigenvalues of $\Phi$, i.e.\  the $a$ period.
Quantum mechanically the poles of the integrand open up into cuts. Contours that wrap around the cuts
still compute the $a$ period, but new contours are available.  They compute the dual periods $a_D$.
As we identified $T(\sigma) = \frac{dt}{d\sigma}$, the contour integrals of $\sigma T(\sigma) d\sigma$ clearly coincide with the
contour integrals of $\lambda$ in the $t$ variable.

To conclude, the use of the resolvent of the bulk theory and the corresponding quantum shifts of the operators $p^{(k)}[u_i]=\mathrm{Tr} \Phi^k$ reflect the global structure associated with the moduli space of vacua.  This information is transmitted to the $2d$ theory through the use of $p^{(k)}[u_i]$ in $\CW$.  This explains the puzzle presented in the introduction about the way the $2d$ theory captures this global structure of the bulk theory.

\subsubsection{Coupling to a chiral doublet}
Now we can go back to the simplest possible GLSM defect: we can couple the bulk gauge theory to a chiral doublet, without any other $2d$ degrees of freedom,
with twisted mass $m$ for the residual $U(1)$ flavor symmetry of the doublet.

First of all, we should make sure this defect is well defined. Classically, one of the two chiral fields becomes massless
at the locus $m^2= u$, and we may worry about having a two-dimensional massless field with non-compact target space.
We will see that quantum mechanically, the chiral fields never become massless.  The classical
logarithmic singularity of the low energy effective twisted superpotential at $m^2= u$ is not present in the quantum theory. More generally, the effective twisted superpotential
\begin{equation}
2 \pi i \CW[m,u] = -\mathrm{Tr}(m + \Phi)\log (m + \Phi)/e
\end{equation}
has the expected analytic properties for a surface defect in pure $SU(2)$ theory.

As the explicit expression for the right hand side in terms of $m$ and $u$ is rather cumbersome, it is convenient to focus on the $m$ and $u$ derivatives.
First, we can look at
\begin{equation} \label{eq:partialmWu}
-2 \pi i \partial_m \CW =  \log \Lambda^2 + \mathrm{arccosh} \frac{m^2- u}{2\Lambda^2} = \log\left(\frac{m^2-u + \sqrt{\left(m^2 - u\right)^2 - 4\Lambda^4}}{2}\right)
~. \end{equation}
This formula is sufficient to verify that the $m$ dependence of the superpotential is physically reasonable.
In any theory with flavor and a twisted mass parameter $m$, the twisted superpotential is defined up to integral multiples of
$m$. This corresponds to the obvious multivaluedness of the $\mathrm{arccosh}$ by multiples of $2 \pi i$.
Modulo $2 \pi i$, the $\mathrm{arccosh}$ function has two branches, which meet at ramification points $m^2 = u \pm 2\Lambda^2$.
These ramification points replace the classical logarithmic singularities at $m^2 \approx u$, much as the monopole and dyon points in the bulk replace the classical W-boson singularity in the pre-potential.

The very existence of two branches of vacua is a very surprising result, which follows mathematically from the multi-valuedness of the generating function $T$.
One of the two branches has a clear semi-classical interpretation for large $m^2- u$, it corresponds to the standard vacuum for the two massive chiral multiplets.
The second branch can only be reached through the cuts between $m^2- u = 2\Lambda^2$ and $m^2- u = -2\Lambda^2$, and appears rather mysterious.
We can gain some small insight into it by looking at the $u$ derivative $\partial_u \CW$, which we can derive by taking a $u$ derivative of $\partial_m \CW$ and integrating over $m$:
\begin{equation}
2 \pi i \partial_u \CW = - \int_m^\infty \frac{dx}{\sqrt{(x^2-u)^2-4\Lambda^4}} ~.
\end{equation}
The integration contour can start at $m$ on either sheet of the square root, depending on which vacuum we are in,
but must be taken to end at infinity on the sheet of the square root which behaves as $x^{-2}$.

Different choices of integration contour  differ by integer linear combinations of $2 \pi i \partial_u a$ and $2 \pi i \partial_u a_D$, as they should.
Indeed, the integrand is nothing else but the holomorphic differential $\frac{dx}{y}$ on the elliptic curve
\begin{equation}
y^2 = (x^2-u)^2-4\Lambda^4 ~.
\end{equation}
The relation
\begin{equation}
\partial_u \CW = \eta \partial_u a + \alpha \partial_u a_D
\end{equation}
tells us that the IR monodromy parameters for the two vacua of the defect coincide with the points on the Jacobian where $x=m$.

In particular, for very large $m$ the semiclassical vacuum has $\alpha = \eta =0$: the chirals are very massive and have a negligible effect on the bulk theory.
On the other hand, for the second vacuum we can compute $\eta =0$ but $\alpha = 1/2$.\footnote{This statement for the result is valid for, say, the standard electric-magnetic frame near the positive real large $u$.} After some careful checks of various normalizations, we see that this monodromy is actually invisible to all the bulk BPS particles, which have even electric charge. In other words, the surface defect in the second vacuum for large
$m$ still has a negligible effect on the bulk degrees of freedom, but it is visible to probes such as a fundamental Wilson loop.
We will see later that the accumulation of magnetic flux in the second vacuum has interesting effects in theories with fundamental flavors.

The existence of the second vacuum is particularly surprising because it holds true at arbitrarily large values of
$u$ and $m^2-u$, where semiclassical considerations usually hold and the $2d$ degrees of freedom are very massive.
The classical supersymmetric Lagrangian couplings of $2d$ chiral fields to the bulk gauge fields relate the
monodromy parameter $\alpha$ to the vev of the moment map for the chiral fields \cite{Gaiotto:2009fs}.
It is hard to understand how massive $2d$ fields may acquire a vev compatible with $\alpha = 1/2$.
We propose a partial solution of this puzzle through the local model described in section \ref{sec:2dZ}.

It is interesting to observe that up to the shift by $\frac{1}{2} a_D$ and $m \log \Lambda^4$ visible in the first derivatives, the effective twisted superpotential
for the second vacuum is identical to the effective twisted superpotential for the semiclassical vacuum of a slightly different UV defect: a chiral doublet of $2d$ charge $-1$. Mathematically, this is just the statement that $T(x)$ has opposite value in the two vacua:
\begin{equation}
T^+(x) + T^-(x) =0
~. \end{equation}

\begin{figure}
\center
\includegraphics[width=3.5in]{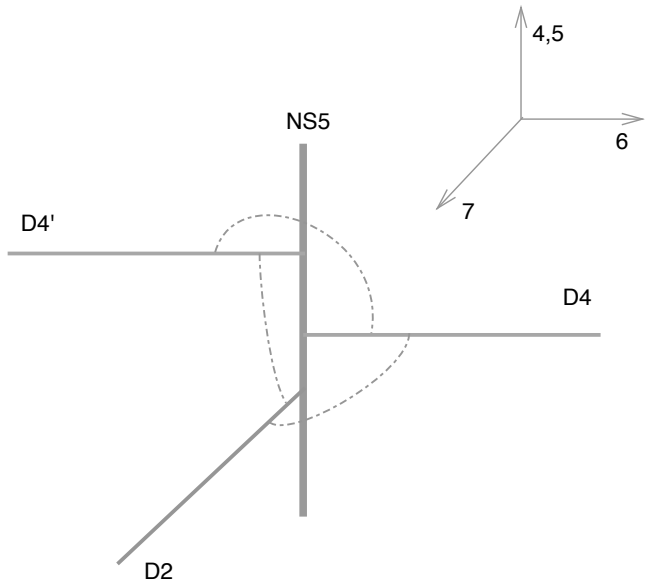}
\caption{The basic ingredients of brane engineering. We depict the directions transverse to the four-dimensional space-time only. A system of D4 branes ending on NS5 branes gives rise to the four-dimensional degrees of freedom. Extra D2 branes ending on the NS5 brane add two-dimensional degrees of freedom. The $4-4'$ strings become $4d$ hypers. The $2-4$ and $2-4'$ strings become $2d$ chirals of opposite $2d$ flavor charge. The three types of fields are coupled by a cubic superpotential.
}
\label{fig:one}
\end{figure}

\begin{figure}
\center
\includegraphics[width=5.5in]{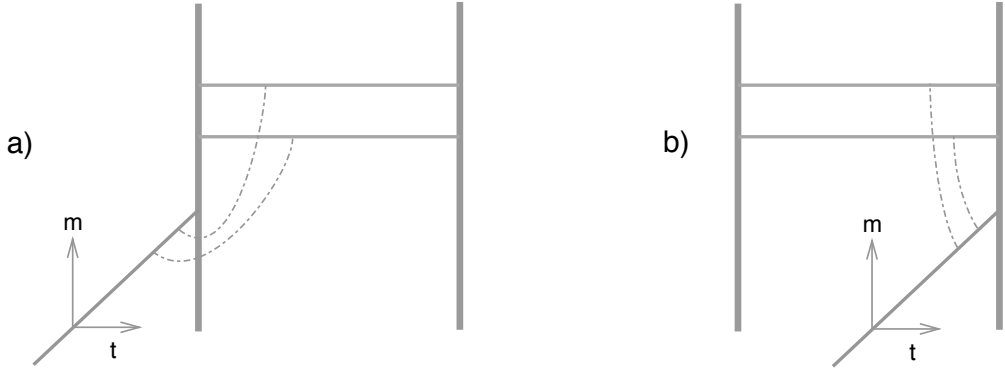}
\caption{The basic surface defect in pure $SU(2)$. The two D4 branes engineer the bulk gauge theory, while the D2 brane ending on the leftmost NS5 brane engineers the chiral doublet.
a) If we end the D2 brane on another NS5 brane far from the system in the $x^7$ direction (not depicted),
we arrive at the full $\mathbb{CP}^1$ GLSM. We indicate the geometric directions corresponding to $m$ (i.e.\  $\sigma$) and
to $t$ (i.e.\  $\partial_m \CW$). Upon lifting to M-theory, the NS5 branes and D4 branes merge into an M5 brane wrapping the curve in the $(\sigma, t)$ coordinates.
The D2 brane becomes an M2 brane ending on the smooth M5 brane
and can move from one NS5 to the other. b) The second branch of $T(\sigma)$ corresponds to the D2 brane ending on the rightmost NS5 brane. The brane construction provides an alternative description in terms of
a chiral doublet of opposite $2d$ flavor charge.
}
\label{fig:two}
\end{figure}

This description of the second branch is natural from the perspective of the brane construction of this system.  It is reviewed, for example in \cite{Alday:2009fs, Gaiotto:2009fs} and is depicted in figures \ref{fig:one} and \ref{fig:two}.
The four-dimensional theory is engineered by a set of D4 branes suspended between two parallel NS5 branes \cite{Witten:1997sc}.
The surface defect is engineered by adding a D2 brane that ends on the system \cite{Hanany:1997vm,Alday:2009fs}.
At large $m$, the D2 brane is far from the D4 branes, and ends on either NS5 brane.
The two choices correspond to the two vacua of the original chiral doublet surface defect. Depending on the choice of
vacuum the D2-D4 open strings produce either a doublet of charge $1$ or a doublet of charge $-1$.


\subsubsection{A few more GLSMs}

We can consider a few variations of this problem. It is pretty clear that we can immediately compute the twisted effective
superpotential for any surface defect defined by coupling the pure $SU(2)$ gauge theory to a doublet of chiral multiplets
inside some general gauged linear sigma model. For example, we could couple $SU(2)$ to the first two chirals in a $\mathbb{CP}^{n+1}$ model: we would start from
\begin{equation}
2 \pi i \CW(\sigma) = t \sigma - \sum_k (\sigma + m_k) \log (\sigma + m_k)-\mathrm{Tr}(\sigma + \Phi)\log (\sigma +\Phi)/e
\end{equation}
and derive the modified twisted chiral ring relation
\begin{equation}
\left(\sigma^2 - u \right) \prod_k (\sigma + m_k) = \Lambda^4 e^{-t}  \prod_k (\sigma + m_k)^2 + e^{t}
~. \end{equation}
Interestingly enough, this reproduces the result of a brane construction described in \cite{Gaiotto:2009fs}.

Another interesting example would be a product gauge theory, $SU(2) \times SU(2)$, with each gauge factor coupled to a separate doublet in a $\mathbb{CP}^3$
gauged linear sigma model. We can start from
\begin{equation}
2 \pi i \CW(\sigma,\Phi, \tilde \Phi) = t \sigma -   \mathrm{Tr}(\sigma -m + \Phi) \log (\sigma -m + \Phi)/e - \mathrm{Tr}(\sigma +m + \tilde \Phi) \log (\sigma +m + \tilde \Phi)/e
~. \end{equation}
The extremum equation becomes
\begin{equation}
\mathrm{arccosh} \frac{(\sigma-m)^2- u}{2\Lambda^2} + \mathrm{arccosh} \frac{(\sigma+m)^2- \tilde u}{2\tilde \Lambda^2} =t-\log \Lambda^2 \tilde \Lambda^2
~. \end{equation}
After some manipulation, this can be brought to the form
\begin{equation}
e^{2 t} -e^t P(\sigma)\tilde P(\sigma) + \left( \Lambda^4 \tilde P(\sigma)^2 +  \tilde \Lambda^4 P(\sigma)^2 -2  \Lambda^4 \tilde \Lambda^4  \right) - e^{-t} \Lambda^4 \tilde \Lambda^4 P(\sigma) \tilde P(\sigma) +  \Lambda^4 \tilde \Lambda^4 e^{-2t} =0
~, \end{equation}
where $P(\sigma) = (\sigma-m)^2- u$ and $\tilde P(\sigma) = (\sigma+m)^2- \tilde u$. Up to minor notation differences,
the same equation was obtained in \cite{Gaiotto:2009fs}, through a brane construction for an $SO(4)$ gauge theory.

Our construction makes manifest that the discontinuities of the effective twisted superpotential $\CW$, which are inherited from the discontinuities of
$\CW(\sigma)$, are integer linear combinations of the periods of the two $SU(2)$ gauge theories. This fact is far from obvious from the twisted chiral ring
equation alone. In this paper, we learn the precise mechanism that allows this, at least for gauged linear sigma models: the coupling to a bulk theory only replaces the one-loop contributions of $2d$ chiral fields with integrated resolvents for the bulk theory, which have discontinuities given by the periods. Extremization of the superpotential preserves the discontinuities.

\subsection{Pure $SU(N)$}

The simplest surface defect which is usually considered in this context arises from the coupling of the pure $SU(N)$ gauge theory to a $\mathbb{CP}^{N-1}$ gauged linear sigma model. It has a simple brane construction, and it gives the standard curve for the theory.
At first, we can simply focus on the coupling of the bulk theory to a set of fundamental chiral fields,
and later add the two-dimensional $U(1)$ gauge field.

\subsubsection{Coupling to a fundamental chiral}
The resolvent of the bulk theory
\begin{equation}
T(x) = \mathrm{Tr}\frac{1}{x + \Phi}= \frac{P'_N(x)}{\sqrt{P_N(x)^2 - 4 \Lambda^{2N}}}
\end{equation}
is expressed in terms of $P_N(x)=\det \left( x+\Phi\right)$.  Integrating it, we derive the $2d$ twisted superpotential $\CW$
\begin{equation} \label{eq:partialmWN}
-2\pi i \partial_m \CW=\mathrm{Tr}\log (m + \Phi) = \log \Lambda^N + \mathrm{arccosh} \frac{P_N(m)}{2\Lambda^N} = \log \frac{P_N + \sqrt{P_N^2 - 4\Lambda^{2N}}}{2} ~.
\end{equation}
This surface defect is similar to the $N=2$ case discussed above. In particular, it has two branches of vacua --
a semiclassical branch with a simple physical interpretation, and a more mysterious second branch.

The monodromy parameters can be read from the integrals of holomorphic differentials on the appropriate hyper-elliptic curve
\begin{equation} \label{eq:partialuWN}
2 \pi i \partial_{u_i} \CW = \int_m^\infty \frac{\partial_{u_i} P_N(x) dx}{\sqrt{P_N(x)^2-4\Lambda^{2N}}}~.
\end{equation}
The monodromy parameters start from $0$ at large $m$ on the semiclassical branch. As we cross into the other branch and go back to large $m$ on the second branch, they reach a
very special value $\alpha_i=\frac{1}{N}$, which is equivalent to zero in a $PSU(N)$ gauge theory, but is non-trivial in an $SU(N)$ gauge theory. In other words, the fundamental chiral defect
at large $m$ on the second branch, is invisible to all excitations of the bulk theory, but is visible as a monodromy of $e^{2 \pi i/N}$ to a fundamental Wilson loop probe. Again, this fact will be important when we discuss theories with fundamental flavors.
The physical origin of the second vacuum should be the same for all $N$, and is partially addressed by the local model in section \ref{sec:2dZ}.

\begin{figure}
\center
\includegraphics[width=6in]{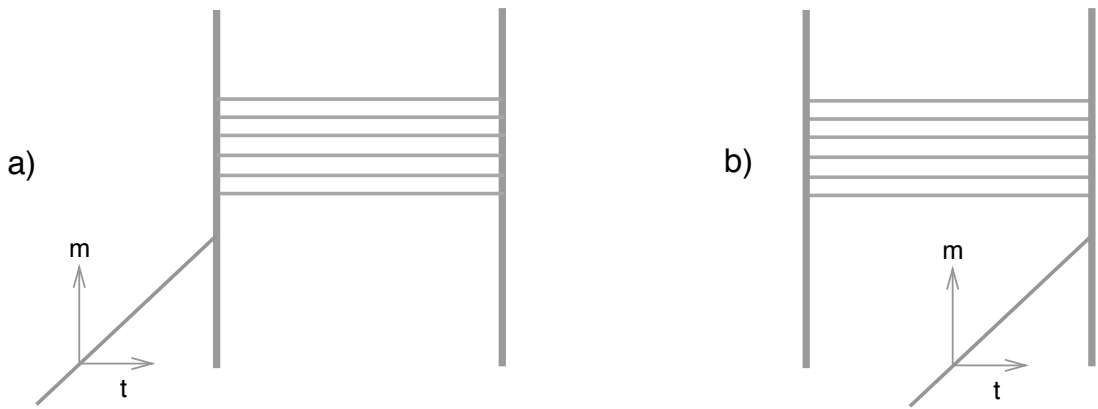}
\caption{The basic surface defect in pure $SU(N)$ (here we depict $N=6$). The two branches of $T(\sigma)$ corresponds to the D2 brane ending on either NS5 brane.
The brane construction provides an alternative description of the second branch in terms of
a chiral anti-fundamental of opposite $2d$ flavor charge as the original chiral fundamental.
}
\label{fig:three}
\end{figure}

Up to the shifts by $m \log \Lambda^{2N}$ and $\frac{1}{N} \sum_i a_D^i$, the effective twisted superpotential
for the second vacuum is identical to the effective twisted superpotential for the semiclassical vacuum of a slightly different UV defect: a chiral anti-fundamental of $2d$ charge $-1$. Again, this is just the statement that $T(x)$ has opposite value in the two vacua:
\begin{equation}
T^+(x) + T^-(x) =0
~. \end{equation}
The brane setup of this system in figure \ref{fig:three} is consistent with this description.

\subsubsection{Coupling to $\mathbb{CP}^{N-1}$}
Now we can couple the previous model to a $2d$ $U(1)$ gauge field.  This leads to the $\mathbb{CP}^{N-1}$ gauged linear sigma model.  Its solution is obtained by a Legendre transform of $\CW$ in (\ref{eq:partialmWN})-(\ref{eq:partialuWN}), and leads to the expected twisted chiral ring relation for the surface defect \cite{Alday:2009fs}:
\begin{equation}
P_N(\sigma) = e^t +\Lambda^{2N} e^{-t}
~. \end{equation}
Equipped with the canonical differential $\lambda = \sigma dt$, this is the standard curve for the theory.

The pure $SU(N)$ surface defect, which is defined by the coupling to a $\mathbb{CP}^{N-1}$ gauged linear sigma model is conjectured to have
a distinct UV description as a GW surface defect.
In general, it is expected to be possible to give an alternative definition of such a defect \cite{Gukov:2006jk,Kanno:2011fw} by starting from the unperturbed bulk theory, and adding
some local $2d$ degrees of freedom to Higgs it to $L$. The degrees of freedom should take the form of a non-linear sigma model with target space $G/L$, coupled
to the bulk fields by gauging the $G$ flavor symmetry acting on the left of the coset.
For $L$ and $G$ of the same rank this is always a K\"ahler manifold, and thus we can take the non-linear sigma model to have $(2,2)$ supersymmetry, as appropriate for a half-BPS surface defect. The K\"ahler parameters for the non-linear sigma model should map to the monodromy parameters.
More precisely, unless $L=G$ the monodromy defects have a moduli space of exactly marginal deformations, corresponding to modifications of the monodromy, which commute with the same
subgroup $L$, complexified by choices of $2d$ theta angles. The dimension of the space equals the number of $U(1)$ factors in $L$.

The simplest surface defect in this class corresponds to the following Levi subgroup of $SU(N)$:
\begin{equation}
L_{1,N-1} = S\left(U(1) \times U(N-1) \right) \,.
\end{equation}
The coset $SU(N)/L_{1,N-1}$ coincides with $\mathbb{CP}^{N-1}$. This fact justifies the conjecture that the coupling of pure $SU(N)$ gauge theory
to a  $\mathbb{CP}^{N-1}$ gauged linear sigma model may produce the same surface defect as the simplest GW surface defect \cite{Gukov:2006jk}.
The conjecture is also supported by the observation that the $U(1)$ part $x$ of $\Phi$ brought to the defect defines a $2d$ twisted chiral operator,
which classically satisfies
\begin{equation}
P_N(x) =0
~. \end{equation}
If we redefine $e^t \to \Lambda^{N}e^{t_{GW}}$ and identify $x$ with $\sigma$, the twisted chiral ring relation
\begin{equation}
P_N(x) = \Lambda^{N}e^{t_{GW}}+\Lambda^{N} e^{-t_{GW}}
\end{equation}
can be thought as arising from ``ramified instanton'' corrections to the classical relation (see \cite{2008arXiv0806.1053K} and references therein for the definition of ramified instantons).

\subsubsection{Coupling to the Grassmanian}
The next simplest surface defect would correspond to the \begin{equation}L_{n,N-n} = S\left(U(n) \times U(N-n) \right) \end{equation} subgroup of $SU(N)$.
The corresponding coset manifold $SU(N)/L_{n,N-n}$ is the Grassmanian, which also admits a gauged linear sigma model description: a $U(n)$ gauge theory coupled to $n \times N$ chiral fields, i.e.\  $N$ fundamentals of $U(n)$, with an obvious overall $SU(N)$ flavor symmetry.
The Grassmanian also admits a dual description in terms of an $U(N-n)$ $2d$ gauge theory.

First, we review the standard solution to the $2d$ gauged linear sigma model \cite{Hanany:1997vm}, and then couple it to the bulk theory.

The twisted effective superpotential is just the sum of contributions from the
eigenvalues $\sigma_i$ of the $U(n)$ gauge multiplet scalars. Correspondingly, one has separate extremum equations
\begin{equation}
P_N(\sigma_i) = e^t
\end{equation}
together with the constraint that the $\sigma_i$ are all distinct roots of that equation
(otherwise the vacuum would not be massive, and the procedure not self-consistent).

We should express these equations in terms of the gauge-invariant twisted chiral operators
$x_a$ of the gauged linear sigma model, which coincide with the symmetric polynomials
of the Coulomb branch scalar fields $\sigma_i$. It is convenient to define a generating function
\begin{equation}
Q(z) = \prod^{n}_i (z - \sigma_i) = \sum_a x_a z^a
\end{equation}

The extremum equations are equivalent to the existence of a factorization
\begin{equation}
Q(z) \tilde Q(z) = P_N(z) - e^t
~. \end{equation}
where $\tilde Q$ is some auxiliary polynomial of degree $N-n$.
Indeed, this guarantees that the $\sigma_i$ are distinct roots of the right hand side.
The gauge-invariant twisted chiral ring relations are obtained
by equating the coefficients of the same powers of $z$ on the left and right hand side of this polynomial equation \cite{Witten:1993xi}.

The coefficients of $\tilde Q(z)$ appear as auxiliary variables in the twisted chiral ring relations.
They could be eliminated, at the cost of a considerable increase in complexity.
On the other hand, they have an obvious interpretation
in terms of the dual description of the Grassmanian theory: they are the gauge-invariant twisted chiral operators
built from the eigenvalues of the dual $U(N-n)$ scalar vevs. The twisted chiral ring relations are written in a duality-symmetric form.

After coupling to the pure bulk $SU(N)$ gauge fields, not much changes.
The twisted chiral ring relation becomes
\begin{equation}
Q(z) \tilde Q(z) = P_N(z) - e^t- \Lambda^{2N} e^{-t}
~. \end{equation}
This is our proposal for the twisted chiral ring relations of the Grassmanian surface defect
and thus for the $L_{n,N-n}$ monodromy surface defect \cite{Gukov:2006jk}.

In order to clarify the relation to the GW surface defect, we can re-define the FI parameter as $e^t \to \Lambda^{N}e^{t_{GW}}$ and write the twisted chiral ring relations as
\begin{equation}
Q(z) \tilde Q(z) = P_N(z) - \Lambda^{N} e^{t_{GW}}- \Lambda^{N} e^{-t_{GW}}
~. \end{equation}

If we bring the non-Abelian bulk scalar $\Phi$ to the $L_{n,N-n}$ GW surface defect, the gauge symmetry breaking
allows us to consider the $U(n)$ and $U(N-n)$ parts of $\Phi$, $\Phi_n$ and $\Phi_{N-n}$,
which behave as twisted chiral operators on the defect. Classically, the vacua of the GW defect are in one-to-one correspondence with the ${N \choose n}$ distinct ways to distribute the eigenvalues of $\Phi$ among $\Phi_n$ and $\Phi_{N-n}$. Indeed, the vev of $\Phi$ must commute with the monodromy at the defect and thus $\Phi$ lies in the unbroken Lie sub algebra.
If we identify the characteristic polynomials of $\Phi_n$ and $\Phi_{N-n}$ with $Q(z)$ and $\tilde Q(z)$,
the classical twisted chiral ring relation becomes
\begin{equation}
Q(z) \tilde Q(z) = P_N(z)
\end{equation}
which is compatible with our quantum-corrected answer.

\subsubsection{Coupling to a flag variety}

A general monodromy defect with $K-1$ moduli for an $SU(N)$ gauge group is associated to the block-diagonal subgroup $L_{(n_k)} = S \left( \prod_k U(n_k) \right)$ with $k=1, \cdots, K$ and $\sum_k n_k =N$.
Based on previous examples, we expect the twisted chiral ring relations to be a quantum deformation
of the classical relation
\begin{equation}
P_N(z) = \prod_k q_k(z)
\end{equation}
where the $q_k(z)$ are monic polynomials of degree $n_k$, whose coefficients are the twisted chiral ring operators
built from the restriction of the bulk scalar field $\Phi$ to the defect. The ${N \choose n_1, n_2,\cdots}$
solutions of the classical relation correspond to all the ways of distributing the eigenvalue of $\Phi$ among the
$U(n_k)$ subgroups of the gauge group. As mentioned before, the vev of $\Phi$ must commute with the monodromy at the defect and thus $\Phi$ lies in the unbroken Lie subalgebra.

Our final answer in equation (\ref{eq:toda}) is indeed of this form. To get there, we need to manipulate carefully the appropriate gauged linear sigma model extremum equations, as we did for the Grassmanian model. First, let's identify the correct two-dimensional matter content. The coset $SU(N)/L_{(n_k)}$ can be given a complex structure and identified with a flag variety,
i.e.\ the space of flags $\IC^{n_1} \subset \IC^{n_1+n_2} \subset \cdots \subset \IC^{N}$. In turn, the flag variety can be given a
2d linear quiver description, with gauge groups $U(n_1) \times U(n_1+n_2) \times \cdots$, bifundamental chirals between each consecutive pairs of nodes, and $N$ fundamentals at the last node. We will denote the ranks of the gauge groups as
\begin{equation}
r_1 = n_1\qquad r_2 = n_1+n_2 \qquad r_3 = n_1+n_2+n_3 \qquad \cdots
~. \end{equation}
Notice that this identification between the coset and the flag manifold uses a choice of order of the $n_k$.
It is natural to wonder how a single GW defect can be associated to a coupling to a variety of inequivalent $2d$ sigma models \cite{Kanno:2011fw}.

If we look back at the simplest $\CP^{N-1}$ example, we can see how this could happen.
The UV $U(1)$ monodromy parameter is supposed to be circle valued classically, but it should also be identified
with the real part of the FI parameter $t$, which is valued on a line. The reason for the discrepancy is that as the UV monodromy parameter approaches
a trivial monodromy, certain ramified instanton corrections become unsuppressed and classical consideration fails. In the pure $SU(N)$ theory the effect is strong enough to push the trivial monodromy point all the way to infinity.

For a general GW surface defect, we may imagine that a similar phenomenon will occur every time we try to make two distinct monodromy eigenvalues close to each other. Thus the classical moduli space of monodromy parameters will decompose into separate quantum moduli spaces, each associated to a distinct cyclic ordering of the monodromy parameters for the defect.
It is conceivable that these distinct phases of the ``same'' GW surface defect in pure $SU(N)$ may
be described by coupling the bulk theory to distinct $2d$ sigma models. Strikingly, our final answer
for the surface defect does indeed depend only on a choice of cyclic order of the $n_k$ rather than
on a choice of linear order as for the $2d$ flag manifold theory in the absence of bulk coupling.

As a first step, we should review the derivation of the twisted chiral ring equations for a flag variety \cite{Givental:1993nc,Kim:1994ny,Astashkevich:1993ks}.
\\ \\
{\it Intermission: the twisted chiral ring of the flag variety GLSM}
\\
Each node of a linear quiver is coupled to a set of fundamental chirals, and to a set of anti-fundamental chirals.
We will use often the characteristic polynomial $Q_k(z)$ of the gauge multiplet scalar at the $k$-th node. The coefficients of
$Q_k(z)$ are the gauge-invariant twisted chiral operators of the $2d$ quiver gauge theory.
If we denote as $t_k$ the FI parameter at the $k$-th node, the same type of analysis as for the Grassmanian gives us
the relation
\begin{equation}
Q_{k+1}(z) - e^{t_k} Q_{k-1}(z) = Q_k(z) q_k(z)
~. \end{equation}
Here we denoted the degree $n_{k+1}$ auxiliary monic polynomial at the $k$-th node as $q_k(z)$. The equation at the first node has $Q_0(z) =1$.
The equation at the last, $(K-1)$-th node has $P_N(z)$ instead of $Q_{K}(z)$.
It is convenient to extend these equations by the definition
\begin{equation}
q_0(z) = Q_1(z)
~. \end{equation}

While in the Grassmanian case the original polynomial $Q$ and the auxiliary polynomial $\tilde Q$
play a parallel role, here the auxiliary polynomials $q_k(z)$ will turn out to be more important than the
original polynomials $Q_k(z)$: we can systematically eliminate the $Q_k(z)$ in favour of the $q_k(z)$,
until we arrive to a single polynomial equation equating $P_N(z)$ to a complicated polynomial in the $q_k(z)$.
Thus we can take the coefficients of the $q_k(z)$ as the basic generators of the twisted chiral ring.
The relation between $P_N(z)$ and the $q_k(z)$ can be expanded into powers of $z$ to get the twisted chiral ring relations
for these generators.\footnote{It is entertaining, but not very useful at this stage, to interpret these equations in terms of continued fractions:
\begin{equation}
\frac{Q_{k+1}}{Q_k} = q_k + \frac{e^{t_k}}{q_{k-1} +\cdots}
~. \end{equation} }

We can rewrite the relations in a compact form in terms of the determinant of a certain matrix.
Indeed, if we define the  $K \times K$ matrix
\begin{equation}
A_{ij}(z) = q_i(z) \delta_{ij} + e^{t_i} \delta_{i, j+1} - \delta_{i, j-1} \qquad i = 0 \cdots K-1 \qquad j = 0 \cdots K-1
\end{equation}
and $Q_0(z)=1$, we can write the equations satisfied by the $Q_i(z)$ as
\begin{equation}
\sum_{j=0}^{K-1} A_{ij}(z) Q_j(z) = \delta_{i, K-1} P_N(z)
~. \end{equation}

Thus if $z$ is a root of $P_N(z)$, it must also be a root of $\det A(z)$. But $\det A(z)$ is a monic polynomial of degree $N$, and thus
we have the relation
\begin{equation}
P_N(z) = \det A(z) ~,
\label{eq:toda}
\end{equation}
which gives the chiral ring of the flag manifold \cite{Givental:1993nc,Kim:1994ny,Astashkevich:1993ks}.

If we consider the most general flag manifold, with all $n_k =1$ and thus all $q_k(z)$ of the form $z - x_k$,
the matrix $A(z)$ can be written as $z - L$, with $L$ being the Lax matrix for an open Toda chain integrable model.
For a more general choice of $n_k$ we can still expand $A$ to a new, bigger matrix of the form $z-L$ with the same determinant.
\\ \\
{\it End of intermission: coupling to the bulk theory}
\\
At first sight, the effect of coupling the $2d$ linear quiver gauge theory to the pure $SU(N)$ bulk theory is quite
ugly. It only affects the equation at the last node, and requires the $2d$ gauge multiplet scalar eigenvalues to be zeroes of the rational function
\begin{equation}
P_N(z) - e^{t_{K-1}} Q_{K-2}(z) - \frac{\Lambda^{2N} e^{-t_{K-1}}}{Q_{K-2}(z)}
~. \end{equation}
With a little bit of work, which we collected in appendix \ref{app:man}, it is possible to re-arrange this
relation to a simpler polynomial form. If we define a new matrix
\begin{equation}
\hat A(z,t) = \begin{pmatrix} q_0(z) & -1 & 0 & \cdots & 0 & 0 & -\Lambda^{2N} e^{-t}\prod_k (- e^{-t_{k}}) \cr
e^{t_1} & q_1(z) & -1& \cdots & 0 & 0 & 0 \cr
\cdots & \cdots & \cdots & \cdots & \cdots & \cdots & \cdots \cr
0 & 0 & 0 &\cdots & e^{t_{K-2}} & q_{K-2}(z) & -1 \cr
- e^t & 0 & 0 & \cdots & 0 & e^{t_{K-1}} & q_{K-1}(z) \end{pmatrix}
\end{equation}
then the twisted chiral ring equations become
\begin{equation}
\det \hat A(z,t) = P_N(z) -e^t - \Lambda^{2N} e^{-t}
~. \end{equation}
The variable $t$ drops out of the final twisted chiral ring equations, but it is useful to include it here.

For the most general case of $n_k =1$, this is the famous result that the curve for pure $SU(N)$ is the spectral curve
for the lax matrix $\hat L$ of a periodic Toda chain integrable system \cite{Gorsky:1995zq,Martinec:1995by,Nekrasov:2003rj}.
We have learned that the periodic Toda chain integrable system can be related to a general GW surface defect, which breaks $SU(N)$ to $U(1)^{N-1}$. Turning on $\Omega$-background along the surface operator turns the classical spectral curve into Baxter equation~\cite{Gadde:2013wq}.

The correspondence between the pure $SU(N)$ gauge theory and the periodic Toda chain integrable system becomes more precise if one compactifies the $4d$ theory on a circle. The moduli space of supersymmetric vacua of any $\CN=2$ theory compactified on a circle is a complex integrable system
whose base is the $4d$ Coulomb branch and whose fiber is parameterized by the choice of electric and magnetic monodromies on the circle.
For pure $SU(N)$ the complex integrable system coincides with the periodic Toda chain.

In general, if we pick any UV surface defect and fix its parameters, the IR data of the defect provides a complex Lagrangian submanifold of the complex integrable system \cite{Alday:2009fs,Gaiotto:2011tf}. The twisted effective superpotential, as a function of the Coulomb branch parameters, is simply the generating function for that Lagrangian submanifold. Indeed, the surface defect in the IR can be described as a GW defect for the Abelian IR theory.
The derivative of the superpotential with respect to the Coulomb branch parameters gives the electric and magnetic monodromies for that defect, which
naturally give a point on the fiber of the complex integrable system for each choice of Coulomb branch parameters.
If we let the UV couplings of the surface defect vary, we will describe a larger submanifold of the complex integrable system, which is not Lagrangian anymore. Rather, the complex symplectic form restricts to the canonical form
\begin{equation}
\Omega = \sum_a dt_a \wedge dx_a ~,
\end{equation}
where $t_a$ are the UV parameters, and $x_a$ the corresponding exactly marginal twisted chiral operators.

Thus our result concerning the generic UV GW surface defect can be stated as follows: as we vary the UV parameters of the surface defect, the IR data for the various vacua
of the theory covers the fiber of the complex integrable system, and thus provides a symplectomorphism between the
space of $(t_a, x_a)$ for the defect and the complex integrable system, which coincides with the periodic Toda chain description.

At least the topological statement about the fiber of the integrable system is perhaps natural, at least at weak coupling: at weak coupling we expect the UV and IR parameters to essentially coincide.
In general, we may expect that for any four-dimensional Lagrangian gauge theory the most general monodromy defect \cite{Gukov:2006jk},
which breaks all gauge groups to the Cartan submanifold, should provide an interesting parameterization of the corresponding  complex integrable
system. It should be possible to make contact with direct instanton calculation such as \cite{Alday:2010vg,Kozcaz:2010yp}.

\subsection{$\CN=2$ SQCD}

In the presence of matter, the resolvent take the form
\begin{equation}
T(\sigma)= \frac{B'(\sigma)}{2 B(\sigma)} + \frac{2 B(\sigma)P'_N(\sigma)-B'(\sigma)P_N(\sigma)}{2 B(\sigma) \sqrt{P_N(\sigma)^2 - 4\Lambda^{2N-N_f}B(\sigma)}}
\end{equation}
with $P_N$ the characteristic polynomial for the Coulomb branch vevs and
\begin{equation}
B(\sigma) = \prod_a (\sigma - m_a)
\end{equation}
for the mass parameters.

Thus we get
\begin{equation}
 \mathrm{Tr}\log (m - \Phi) = \log \frac{P_N(m) + \sqrt{P_N(m)^2 - 4\Lambda^{2N-N_f}B(m)}}{2}
\end{equation}
and learn that the basic defect defined through coupling to a set of $2d$ fundamental chiral fields still has a semiclassical vacuum and a second, hidden
vacuum. The main difference is that the second branch of vacua has interesting new logarithmic singularities at the locations of the bulk mass parameters,
$m = m_a$, which signal the presence of extra $2d$ chiral fields, which carry no gauge charge, $2d$ flavor charge $1$ and transform in the fundamental representation of the bulk flavor group.
These $2d$ fields may be the manifestation of some sort of Landau levels for the bulk fundamental matter, localized near the defect by the magnetic flux on the second branch.

There is a useful way to express the relation between the two branches of the $T(x)$ function
\begin{equation}
T^+(x) + T^-(x) = \frac{B'(\sigma)}{B(\sigma)} ~.
\end{equation}
It can be interpreted as suggesting that the surface defect has a second UV description, which makes the second branch semiclassical: a set of $2d$ chiral fields in the anti-fundamental representation of the gauge group,
with $2d$ flavor charge $-1$, together with the aforementioned set of $2d$ chirals in the fundamental representation of the flavor group.

\begin{figure}
\center
\includegraphics[width=6in]{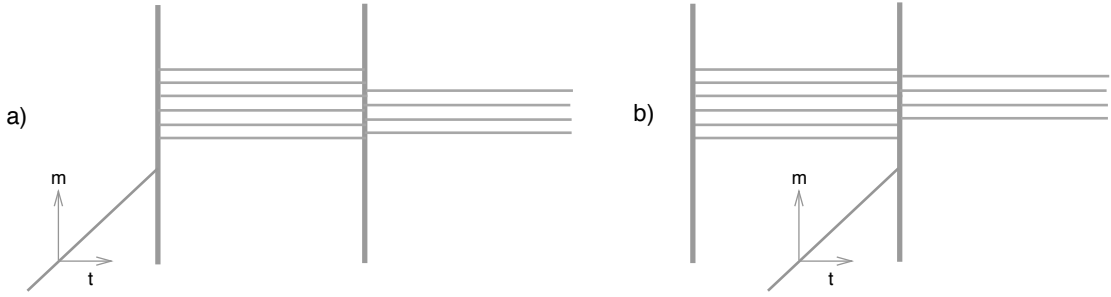}
\caption{The basic surface defect in $SU(6)$ $N_f=4$ SQCD. The extra semi-infinite D4 branes on the right add the bulk flavors to the gauge theory. a)
If the D2 brane ends on the leftmost NS5 brane, we have the usual set of $2d$ fundamental chirals of $2d$ flavor charge $1$.
 b) The second branch of $T(\sigma)$ corresponds to the D2 brane ending on the rightmost NS5 brane. The brane construction provides an alternative description in terms of
a chiral $SU(6)$ anti-fundamental of $2d$ flavor charge $-1$, a chiral fundamental of the $U(4)$ flavor group and a superpotential coupling to the bulk hypermultiplets.
}
\label{fig:four}
\end{figure}

This is natural in the brane constructions
of the $\mathbb{CP}^{N-1}$ defect, reviewed for example in \cite{Alday:2009fs,Gaiotto:2009fs} and depicted in figure \ref{fig:four}.
The brane description supports an alternative description of the second branch of vacua and
suggests the existence of an ordinary (not twisted) superpotential coupling involving the product of these two types of $2d$ fields and a
bulk hypermultiplet with appropriate charges:
\begin{equation}
\int d^2 \theta d^2 x q_\alpha \tilde q^A M^\alpha_A|_{z=0}
~. \end{equation}
Here $q_\alpha$ ($\alpha = 1,...,N_c$) are the $2d$ chirals in the anti-fundamental representation of the gauge group, $\tilde q^A$ ($A = 1,...,N_f$) are the
$2d$ chirals in the fundamental representation of the $4d$ flavor group, and $M^\alpha_A|_{z=0}$ are components of the bulk hypermultiplets restricted to the defect.

Taking inspiration from the brane constructions, there is a natural generalization of these defects, which will be important when we look at quiver gauge theories.
We split the fundamental hypermultiplets into two groups -- one with $N_f^+$ hypermultiplets and the other with $N_f^-$ hypermultiplets ($N_f=N_f^+ + N_f^- $).  Correspondingly, we split the characteristic polynomial $B(x) = B_+(x) B_-(x)$ for the $U(N_f^+) \times U(N_f^-) \subset U(N_f=N_f^+ + N_f^-)$ of the flavor group.  Now we consider the usual $2d$ chirals in a fundamental representation of the gauge group and $2d$ flavor charge $1$,
add $2d$ chirals in an anti-fundamental representation of
the flavor subgroup $U(N_f^+)$ and $2d$ flavor charge $-1$. We can then add an ordinary superpotential coupling them to
the first group of hypermultiplets
\begin{equation}
\int d^2 \theta d^2 x q^\alpha \tilde q_A \tilde M^\alpha_A|_{z=0}
~. \end{equation}
Here $\tilde M$ is the other complex scalar field in the bulk hypermultiplet.

\begin{figure}
\center
\includegraphics[width=6in]{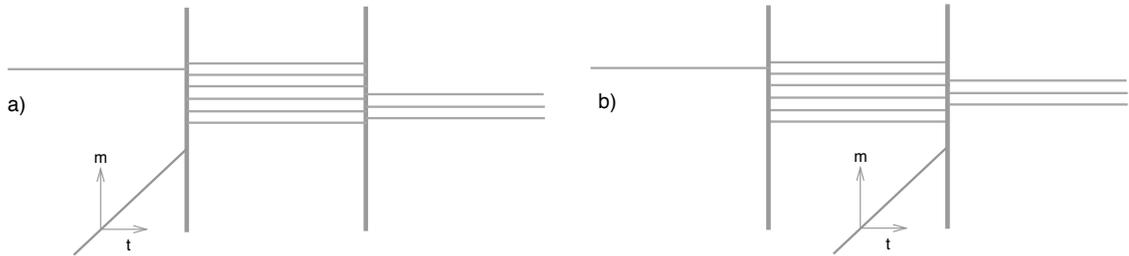}
\caption{Another simple surface defect in $SU(6)$ $N_f=4$ SQCD. Moving the semi-infinite D4 brane from the right to the left does not change the bulk theory. It adds an extra $2d$ chiral to the description of the basic
surface defect a), and removes it from the dual description on the second branch b).
}
\label{fig:five}
\end{figure}

The effective twisted superpotential is modified accordingly by the contribution of the new $2d$ fields, to
\begin{align}
- 2\pi i \partial_m \CW &= \log \frac{P_N(m) + \sqrt{P_N(m)^2 - 4\Lambda^{2N-N_f}B(m)}}{2 B_+(m)} \cr &= \log \frac{2 \Lambda^{2N-N_f} B_-(m)}{P_N(m) - \sqrt{P_N(m)^2 - 4\Lambda^{2N-N_f}B(m)}} ~.
\end{align}
Looking at the second expression or at the brane picture in figure \ref{fig:five}, on the second branch
we expect a dual description of the surface defect,
in terms of $2d$ chirals in an anti-fundamental representation of the gauge group and $2d$ flavor charge $-1$ and $2d$ chirals in an fundamental representation of
the flavor subgroup $U(N_f^-)$ and $2d$ flavor charge $1$, with appropriate ordinary superpotential couplings to the bulk hypermultiplets.

Starting from these elementary defects and gauging $2d$ flavor symmetries we can compute the IR description for a wide range of surface defects.
The calculations add little to the examples we have already encountered, and thus we prefer to move on to more intricate bulk theories.

\subsection{Four dimensional quiver gauge theories}
Now we would like to briefly discuss surface defects defined in $\CN=2$ quiver gauge theories
with unitary gauge groups, in order to make contact with the beautiful results of \cite{Nekrasov:2012xe}.
In order to facilitate the comparison, it is useful to add one more notation. Given a resolvent $T(z)$ for any $U(N)$ gauge group, we will define
\begin{equation}
Y(x) = \exp \int T(x) = \exp \mathrm{Tr} \log (x-\Phi)
~. \end{equation}
On the semiclassical sheet, $Y(x)$ is close to the characteristic polynomial $P_N(x)$. In a quiver gauge theory,
one has a resolvent and a $Y_a(x)$ function for any node of the quiver.

Although the specific expression for $Y_a(z)$ depends on the full gauge theory structure,
it satisfies a basic property derived in \cite{Nekrasov:2012xe}, relating the values across the cuts in the
semiclassical sheet:
\begin{equation} \label{eq:Ysys}
Y_a^+(x) Y_a^-(x) = B_a(x) \prod_{b} Y_b(x + m_{ab}) ~,
\end{equation}
where $B_a(x)$ is the characteristic polynomial for the fundamental hypers at the node, multiplied by the appropriate power of the gauge coupling scale at that node,
the index $b$ runs over the nodes contiguous to $a$ and $m_{ab} = - m_{ba}$ is the mass for the bifundamental hypers.

We can give a direct interpretation of this equation in terms of surface defects.
Consider a defect defined in terms of a set of fundamental $2d$ chirals at the $a$-th node. The effective superpotential satisfies
\begin{equation}
-2 \pi i\partial_m \CW = \log Y_a(m)
~. \end{equation}
The naive vacuum for this defect corresponds to the value of $Y_a(m)$ on the first sheet.
If we go across the cuts to the second sheet, we find
\begin{equation}
-2 \pi i\partial_m \CW = \log M_a(m) + \sum_b \log Y_b(x + m_{ab}) -\log Y_a(m)
~. \end{equation}

This means that although we have modified the bulk theory rather drastically, adding new gauge fields, the basic physics of the
second branch of vacua has not changed from the SQCD case: it is still true that we can find a second UV description where the second branch of vacua has a semi-classical meaning.
The second UV description involves exactly the same degrees of freedom that we encountered in SQCD: $2d$ gauge anti-fundamentals of $2d$ flavor charge $-1$ and $2d$ chirals of $2d$ flavor charge $1$
associated to all the bulk hypers coupled to the $a$-th node, i.e.\
a set carrying a fundamental representation of the residual flavor group at the $a$-th node and
a set for each contiguous node, carrying a fundamental representation of the contiguous node's gauge group.

The main effect of having a quiver gauge theory is that now the functions $Y_b(x + m_{ab})$ associated to the contiguous nodes
have new cuts, across which we find new sheets that can be given a semi-classical description by replacing the $2d$ chirals
carrying a fundamental representation of the contiguous nodes with new sets of chirals, according to (\ref{eq:Ysys}), etc.
At least for asymptotically free or conformal quiver gauge theories,
all these new sheets glue together nicely into a curve for the bulk theory \cite{Nekrasov:2012xe}.

Linear quiver gauge theories are a particularly neat example of this structure.
For simplicity, we can assume the absence of bulk fundamental matter away from the quiver's ends. This limitation can be easily removed.
A defect at the first node, with superpotential
\begin{equation}
-2 \pi i\partial_m \CW = -\log M_1(m) +\log Y_1(m)
\end{equation}
on the first sheet will have
\begin{equation}
-2 \pi i\partial_m \CW = \log Y_2(m+m_{12}) -\log Y_1(m)
\end{equation}
on the second sheet.
Then we can go to the next sheet for $Y_2$, and find
\begin{equation}
-2 \pi i\partial_m \CW = \log Y_3(m+m_{12}+m_{23}) -\log Y_2(m+m_{12})
\end{equation}
and so on and so forth.

\begin{figure}
\center
\includegraphics[width=6in]{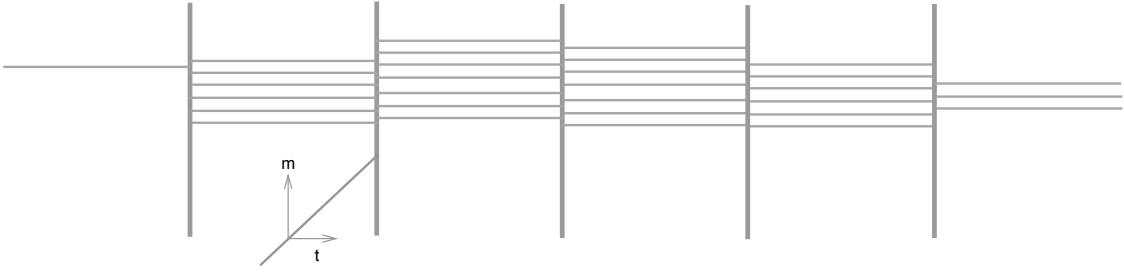}
\caption{A simple surface defect in a quiver gauge theory with gauge group $SU(6) \times SU(7) \times SU(7) \times SU(6)$.
The five branches of $T(\sigma)$ can be described by the D2 brane ending on the five NS5 branes in the system.
}
\label{fig:six}
\end{figure}

If we gauge the $U(1)$ flavor symmetry, we have equations such as
\begin{equation}
e^t = \frac{Y_1(\sigma) }{M_1(\sigma)}
~. \end{equation}
From the localization calculation or the brane solution depicted in figure \ref{fig:six}, we know that $Y_1(\sigma)$ must be such that
$e^t$ is a solution of
\begin{equation}
e^t M_1(\sigma) + P_1(\sigma) + c_2 P_2(\sigma) e^{-t} + c_3 P_2(\sigma) e^{-2 t} \cdots =0 ~,
\end{equation}
where $P_a(\sigma)$ is the characteristic polynomial for the $a$-th gauge group.
In other words, for a linear quiver with $L$ gauge groups, $Y_1(x)/M_1(x)$ is the solution of a degree $L+1$ polynomial equation.
The other solutions of the same equation coincide with the $Y_{i+1}/Y_i$.
From the point of view of the brane system, the sheets of $Y_1(x)/M_1(x)$ correspond to the various NS5 branes the D2 brane can end on.

Using these relations, together with the explicit solution for $Y(x)$ \cite{Nekrasov:2012xe}, we can
solve for the IR dynamics of very general surface defects in the four-dimensional linear gauge theories.
We will focus now on a theory which provides a very special example of this structure.

\subsection{The $\CN=2^*$ theory}
Circular quiver gauge theories, or more simply the $\CN=2^*$ $SU(N)$ gauge theory,
have a rather complicated resolvent, which involve the inversion of elliptic functions.
The basic relation from localization becomes
\begin{equation}
Y^+(x) Y^-(x) = q Y(x + \frak{m}) Y(x - \frak{m}) \qquad, \qquad  q = e^{2 \pi i \tau}
~. \end{equation}
We denote as $\frak{m}$ the bulk mass parameter.

It is clear from \cite{Nekrasov:2003rj} that it is hard to work with $Y(x)$.
It is simpler to work with the ratio
\begin{equation}
\tilde Y(m) =  \frac{Y(m+\frak{m}/2)}{Y(m-\frak{m}/2)}
~. \end{equation}
Indeed, it is possible to fix $\tilde Y$ from its analytic properties.
If we use the basic relation to analytically continue $\tilde Y$ across the cuts of the numerator, we get that on the next sheet
\begin{equation}
\tilde Y(m) = q \frac{Y(m+3\frak{m}/2)}{Y(m+\frak{m}/2)}
~. \end{equation}
etcetera.

\begin{figure}
\center
\includegraphics[width=4in]{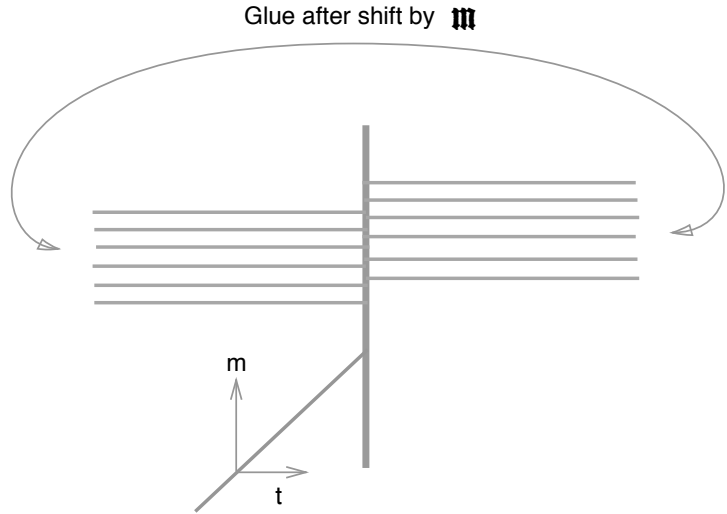}
\caption{A simple surface defect in the $\CN=2^*$ $SU(6)$ gauge theory. In order to add the $\CN=2^*$ mass deformation, the $x^6$ direction is compactified on a circle,
and the $x^4 + i x^5$ directions are fibered on the circle in such a way that they shift by $\frak{m}$ as $x^6$ shifts by one period.
The brane system produces the $\CN=2^*$ mass deformation of a defect with $(4,4)$ SUSY.
}
\label{fig:seven}
\end{figure}

The standard strategy is to consider an affine fibration on an elliptic curve, i.e.
a space with coordinates $(\sigma, t)$ identified under $t \to t+2 \pi$ and
under $(\sigma \to \sigma + \frak{m}, t \to t + 2 \pi \tau)$. The equation
\begin{equation}
e^t = \tilde Y(\sigma)
\end{equation}
defines a nice curve in the affine fibration, a somewhat complicated deformation of $P_N(\sigma)=0$:
\begin{equation}
\theta_1(\tau, t )P_N(\sigma) + \frak{m} \partial_t \theta_1(\tau, t )\partial_\sigma P_N(\sigma) + \cdots + \frac{\frak{m}^k}{k!} \partial_t^k \theta_1(\tau, t )\partial_\sigma^k P_N(\sigma) + \cdots=0
~. \end{equation}
This is the standard Seiberg-Witten curve for the theory, which can be derived from M-theory by the setup in figure \ref{fig:seven} and coincides with the spectral curve for an elliptic Calogero-Moser integrable system \cite{Witten:1997sc,Nekrasov:2003rj,Nekrasov:2012xe}.

These facts suggest that we will have an easier time working with surface defects which involve $\tilde Y(m)$
rather than $Y(m)$. This means coupling the bulk theory
to a set of $2d$ fundamental chirals together with a set of anti-fundamental chirals,
all of charge $1/2$ under the bulk flavor symmetry and opposite charge under a $2d$ flavor symmetry.
These chirals can be coupled by an ordinary superpotential to the bulk adjoint hypermultiplet.

Supersymmetry supports this idea: this defect is simply the
$\CN=2^*$ mass deformation of the simplest half-BPS defects in $\CN=4$ $SU(N)$ gauge theory,
where the bulk theory is coupled to a $2d$ hypermultiplet in a fundamental representation, so to preserve
$(4,4)$ supersymmetry. As the bulk SUSY is broken to $(2,2)$
by the bulk $\CN=2^*$ mass deformation, the superpotential couplings force one to add mass $\frak{m}/2$ to the chiral fields in $2d$ hypers
(and $-\frak{m}$ to the chiral fields in $2d$ vectormultiplets for more general $(4,4)$ gauged linear sigma models).

The effective twisted superpotential for this basic defect satisfies
 \begin{equation}
-2 \pi i\partial_m \CW = \log \tilde Y(m)
~. \end{equation}

If we gauge the $U(1)$ $2d$ flavor symmetry, we get an extremum equation
\begin{equation}
e^t = \tilde Y(\sigma) ~,
\end{equation}
which can be inverted to recover the standard curve.

With sufficient patience we can use these facts to solve for the IR dynamics of (the $\CN=2^*$ mass deformation of) any half-BPS defect in $\CN=4$ $SU(N)$ gauge theory,
which admits a $(4,4)$ $2d$ gauged linear sigma model description. Since $2d$ matter fields are in hypermultiplets,
only the ratio $\tilde Y(m)$ enters the equations, and one may attempt to do calculations similar to these we did for the pure $SU(N)$ theory.
In particular, coupling to the $(4,4)$ $2d$ linear quiver gauge theories, which realize the cotangent bundle to flag manifolds should reproduce the $\CN=2^*$ mass deformation
of the original half-BPS GW surface defects.

It would be interesting to verify that the general GW surface defect directly reproduces the elliptic Calogero-Moser Lax matrix, similarly to what happened with the periodic Toda chain in pure $SU(N)$.
Notice that solving for the effective twisted superpotential for the $(4,4)$ $2d$ linear quiver gauge theories themselves, before coupling to bulk degrees of freedom,
is known to be equivalent to the problem of diagonalizing the trigonometric Calogero Moser
Lax matrix (see \cite{Gaiotto:2013bwa} and references therein). The extremum equations
 take the form of Bethe equations for an XXX spin chain \cite{Nekrasov:2009uh}. The coupling to the $4d$ bulk theory deforms the XXX Bethe equations by
replacing the phase $\frac{P_N(m+\frak{m}/2)}{P_N(m-\frak{m}/2)}$ acquired by the propagation of the basic Bethe roots around the chain
by $\frac{Y(m+\frak{m}/2)}{Y(m-\frak{m}/2)}$.
It would be nice to verify that this modification truly deforms trigonometric Calogero-Moser to elliptic Calogero-Moser.

\section{Local models} \label{sec:BPS}
Until now, we have concerned ourselves with the infrared behavior of surface defects at general points in the Coulomb branch of the bulk theory
and/or of the twisted masses for $2d$ flavor symmetries. At such generic points, the surface defect is well described as a monodromy defect \cite{Gukov:2006jk} for the IR gauge fields.
At special points in the Coulomb branch, singularities in the infrared data signal the appearance of
light BPS particles and an infrared description should include extra degrees of freedom.
We will now describe various examples of this phenomenon.

\subsection{A light $2d$ chiral particle} \label{sec:2dZ}
Consider, as an example, the defect defined by the chiral doublet in the pure $SU(2)$ gauge theory.
If we make the $2d$ twisted mass $m$ very large, the two-dimensional dynamics can only become interesting
if we also tune the Coulomb branch parameter $u$ to be large, close to the point $u \approx m^2$. Since $u$ is large we can approximate it $u \approx a^2$.  Then, the
classical masses for the $2d$ particles are $m \pm \sqrt{u} \approx m \pm a$. Expanding around $m \approx a$
the light degrees of freedom consist of a bulk Abelian gauge field coupled to a single $2d$ chiral field of charge $1$ and mass $m-a$.

Naively, such a theory would have a very simple effective twisted superpotential
\begin{equation}
2 \pi i \CW_0 = -(m-a) \log (m-a)/e ~,
\end{equation}
with a single vacuum and a logarithmic singularity at $m-a=0$.
On the other hand, taking the same scaling limit in the exact solution for the non-Abelian model (\ref{eq:partialmWu}) we find
\begin{equation} \label{eq:partialmWa}
-2 \pi i \partial_m \CW_1 \approx  \log \Lambda^2 + \mathrm{arccosh} \left(\frac{m-a}{\Lambda^2/a}\right) = \log\left(m-a + \sqrt{(m-a)^2 - \Lambda^4/a^2}\right) +\log a ~,
\end{equation}
which has two vacua, and replaces the logarithmic singularity with two branch points.
The non-Abelian degrees of freedom, albeit very massive, manage to deeply affect the physics of the two-dimensional chiral field.  A low energy description of this effect of the massive modes should be describable in terms of a high dimension operator correction to the theory of a single free chiral superfield $Z$.

We will find it useful to look at a mirror description of the problem. A single $2d$ free chiral field $Z$ has a well known mirror description
\cite{Hori:2000kt} in terms of a twisted chiral field $Y$ with a twisted superpotential
\begin{equation} \label{eq:naiveY}
2 \pi i \CW_{HV} = (m-a)Y + e^{-Y}
~. \end{equation}
Integrating out $Y$ gives back the standard one-loop result $\CW_0$. The field $Y$ is classically related by T-duality to a
chiral field $w= \log Z$ valued in a cylinder. But since the point $Z=0$ is at finite distance, there is no winding symmetry associated with $w$.  Correspondingly, there should not be a shift symmetry for $Y$.  The extra $e^{-Y}$ interaction breaks this shift symmetry.

The theory with (\ref{eq:naiveY}) is equivalent to the free theory of $Z$.  But our local model is incomplete and could be corrected by higher dimension operators.  This is not surprising.  It is common for a local description to be deformed by such higher dimension operators.  For example the simple Abelian theory near the monopole point in the bulk is corrected by high dimension operators suppressed by powers of $\frac{1}{\Lambda}$.  Such corrections are not universal.  They depend on the specific UV completion of the theory.  The same is true for our local model.  Its twisted superpotential can be corrected at large $|Y|$.

Specifically, we claim that there is a simple correction term and we should study
\begin{equation}
2 \pi i \CW_{new} = (m-a)Y + e^{-Y} - \eta^2 e^Y
~. \end{equation}
Integrating out $Y$, we find
\begin{equation}
(m-a)= e^{-Y} + \eta^2 e^Y ~;
\end{equation}
i.e.\
\begin{equation}
Y = \log\left(\frac{m-a - \sqrt{(m-a)^2 - 4 \eta^2}}{2 \eta^2}\right)
\end{equation}
and
\begin{equation}
-2 \pi i \partial_m\CW_{new} = \log\left(\frac{m-a + \sqrt{(m-a)^2 - 4\eta^2}}{2}\right)
~. \end{equation}
It matches the non-Abelian result (\ref{eq:partialmWa}) if $\eta^2 = \frac{\Lambda^4}{4a^2}$ and we add to it the trivial contribution of the massive particle $-2\pi i \partial_m\CW |_{massive}=\log 2a$.

We can try to give a physical interpretation to the $\eta^2 e^Y$ correction term. The $Y$-dependence suggests that some bulk effect
allows the $2d$ field $Z$ to have BPS vortex configurations with an unusual negative vortex charge.
The overall coefficient suggests a bulk one-instanton effect. In an appropriate sense, a bulk instanton is
related to a modification of the Abelian gauge bundle at the location of the instanton. If we bring the bulk instanton
to the surface defect, the modification should allow the $2d$ chiral field to have a pole at the location of the instanton,
as required for a BPS vortex of negative charge.

In other words, the target space of the $Z$ theory has the topology of a sphere, with fixed points at $0$ and $\infty$ under the rotation symmetry. When the twisted mass is turned on, we find a vacuum at each of these points. The vacuum at $0$ is the semiclassical one and small fluctuations around the vacuum give a massive BPS particle of charge $1$.
The vacuum at $\infty$ is the hidden one, and small fluctuations around this vacuum give a massive BPS particle of charge $-1$. This is the same as we gleaned from the solution of the $SU(2)$ and $SU(N)$ complete models.

We are left with the problem of understanding the two branch point singularities which replace the naive logarithmic singularity
associated to a light $2d$ particle. The disappearance of the $2d$ chiral singularity is somewhat analogous to the disappearance
of the W-boson singularity in the Coulomb branch: the $2d$ chiral multiplet ceases to exist as a BPS particle
and decays to new BPS particles associated to the branch point singularities. It is useful to review briefly
the general properties of the $2d$ BPS particles that live on surface defects.

\subsection{BPS particles and solitons}

As far as the bulk theory is concerned, the typical codimension one singularity in the Coulomb branch is associated with a single BPS hypermultiplet becoming massless.
In some electric-magnetic duality frame, where the BPS hypermultiplet is electrically charged, with central charge $a$, the dual period
$a_D$ has a logarithmic singularity:
\begin{equation}
a_D \sim \frac{a}{2 \pi i} \log a + \cdots
\end{equation}
and a monodromy
\begin{equation}
a_D \to a_D + a
\end{equation}
around the singular locus $a=0$.

The close relation between the analytic structure of the periods and the BPS spectrum of the bulk theory was used
to determine the BPS spectrum of simple $\CN=2$ theories starting with \cite{Seiberg:1994rs}. Careful considerations based on the Kontsevich-Soibelman wall-crossing formula \cite{2008arXiv0811.2435K,Gaiotto:2008cd} and on the notion of framed BPS degeneracies \cite{Gaiotto:2010be} make the correspondence rather precise. In particular, they provide a systematic, if cumbersome,
algorithm allowing to determine the full BPS spectrum on the whole Coulomb branch from the knowledge of the light BPS spectrum near the complex codimension
one singularities \cite{Gaiotto:2012rg}. Generically, the codimension one singularities are all of the basic type associated to a single BPS hypermultiplet and thus the spectrum is uniquely determined.

The same considerations apply to surface defects. Surface defects support an interesting spectrum of $2d$ BPS excitations.  These includes both BPS ``particles''
in a given vacuum of the surface defect, and BPS ``solitons'' that interpolate between two vacua of the defect \cite{Gaiotto:2011tf}. The central charges of BPS particles
are integer linear combinations of the periods of the bulk theory and possibly extra $2d$ twisted masses. Indeed, they carry well defined, integral gauge and flavor charges.
The central charge of a BPS soliton between two vacua ``$i$'' and ``$j$'' of the surface defect receives an extra contribution given by the difference in twisted effective superpotential between the two vacua, i.e.\
\begin{equation}
\CZ = \CW_i - \CW_j + q_e \cdot a + q_m \cdot a_D + q_f \cdot m
~. \end{equation}
Notice that the twisted effective superpotentials are multivalued functions, which can shift by integer linear combinations of the periods and twisted masses.
Correspondingly, the gauge and flavor charges assigned to a BPS soliton undergo affine-linear monodromies
around singularities of the Coulomb branch.

Again, considerations based on the $2d$-$4d$ wall-crossing formula and on the notion of framed BPS degeneracies provide a systematic algorithm
to determine the full $2d$ BPS spectrum from the knowledge of the light $2d$ BPS spectrum near the complex codimension
one singularities of the Coulomb branch \cite{Gaiotto:2012rg}. Generically, we have two types of codimension one singularities: logarithmic singularities
and branch loci. Logarithmic singularities are associated to either light $2d$ or $4d$ BPS particles, and correspond to a divergence of
the form
\begin{equation}
\CW_j \sim \omega_j \frac{a}{2 \pi i} \log a + \cdots
\end{equation}
for some integers $\omega_j$. Notice that if $4d$ particles of mass $a$ are present, the values of the integers $\omega_j$ depend on the choice of branch for $\CW_j$,
and we can often find a branch where $\omega_j=0$.

Branch loci are associated with light $2d$ BPS solitons and correspond to the branch locus of the fibration of the space of vacua $\CV$ over the Coulomb branch $\CB$.
The local model for a branch locus \cite{Gaiotto:2011tf} is a $2d$ twisted Landau-Ginzburg theory with a cubic twisted superpotential
\begin{equation}
\CW = \frac{\sigma^3}{3} - \delta u \sigma
\end{equation}
twisted chiral ring relation
\begin{equation}
\sigma^2 = \delta u
\end{equation}
and low energy effective twisted superpotential
\begin{equation}
\CW = -\frac{2}{3} \delta u^{3/2}
~. \end{equation}
This theory has a single BPS soliton between its two vacua.
The power-law exponent of $3/2$ in the singular part of $\CW$ near a branch point is expected to be generic: the effective GW parameters and thus $\partial_u \CW$ should be regular at the branch locus.

The simplest example of the chiral doublet coupled to the pure $SU(2)$ theory provides a neat manifestation of these phenomena.
The only logarithmic singularities appear at the standard monopole and dyon points of the Coulomb branch.
In particular, this shows that the chiral doublet BPS particles, much as the W-bosons, decay before they can become massless.
At the branch loci $u = m^2 \pm 2 \Lambda^2$ we find the expected behavior of the superpotential.
For example at $u \sim m^2 + 2 \Lambda^2$ we find, up to regular terms,
\begin{equation}
\CW \sim \frac{1}{3 m \Lambda} (u - m^2 - 2 \Lambda^2)^{3/2}
~. \end{equation}

In order to encounter $2d$ particle singularities, we can look at the surface defect defined by a $2d$ chiral doublet
coupled to the $SU(2)$ theory with flavor. On the non-classical branch of vacua we have singularities
whenever $m$ equals the masses $m_a$ of the bulk hypermultiplets, due to the corresponding poles in $T(x)$.
The singularities correspond to $2d$ BPS particles with pure $2d$ and $4d$ flavor charge
\begin{equation}
2 \pi i \partial_m \CW = - \log B(m)
~. \end{equation}
Note that these singularities appear as functions of the masses, and not at special loci on the Coulomb branch.
We are not aware of any example of $\CN=2$ field theory and surface defect where a $2d$ BPS particle with bulk gauge charges becomes massless
in the absence of a massless $4d$ BPS particle with the same charge. Presumably, the bulk instanton effects
always lift the naive logarithmic singularities.

\subsection{Collision of singularities}
The infrared physics can become much more interesting at special values of the parameters, where two or more basic codimension one singularities collide. The most famous example are Argyres-Douglas theories \cite{Argyres:1995jj}, strongly interacting superconformal fixed points, which provide the infrared description
of $\CN=2$ theories near a locus where mutually non-local BPS particles (say a monopole and an electron) become simultaneously light \cite{Argyres:1995xn}.
In the presence of surface defects, the simplest non-trivial setup arises from the collision of a branch locus and a logarithmic singularity in the Coulomb branch. The collision of several branch loci away from logarithmic singularities can also be interesting, but it is typically described by some purely two-dimensional twisted Landau-Ginzburg theory.

The simplest local model for the collision of a branch locus and a logarithmic singularity was described in \cite{Gaiotto:2011tf}.
We can derive it readily from a scaling limit of either the chiral doublet surface defect or the $\mathbb{CP}^1$ defect in the pure $SU(2)$ gauge theory. In the latter, we can scale
\begin{equation}
u=-2 \Lambda^2 + \Lambda a \qquad \sigma = \Lambda^{1/2} x \qquad e^t = \Lambda^2 \left(1+  \Lambda^{-1/2} z\right)
~. \end{equation}
We find
\begin{equation}
2 \pi i \partial_z \CW = x  \qquad  x^2 = z^2 + a
~. \end{equation}

We can fix the $a$ dependence by comparing with the $a \to 0$ limit of $\partial_u \CW$.
\begin{equation}
2 \pi i \CW \sim -z^2/2 -a/2 \log \frac{z}{2\Lambda^{1/2}}+ \int_z^{\infty} \left[ x(z') - z' - \frac{a}{2z'} \right]dz'
~. \end{equation}
A very similar expression would arise from scaling the doublet defect, with $z$ replaced by a scaled $m$.

We have enough information to start building a physical picture of the system near the monopole point. Notice that we went to an electric-magnetic duality frame,
where the light bulk hypermultiplet is electrically charged, with mass $a$.
At large $z\sim \sqrt{\Lambda}$ on the first branch of vacua, away from the branch locus, the surface defect has a very small effect on the physics of the light bulk hyper.
If we read off the defect's GW monodromy parameters from
\begin{equation} \partial_a\CW = \eta + \tau \alpha
\end{equation}
we find a very small monodromy parameter, as the left hand side is of order $\log \frac{z^2}{\Lambda}$ while $\tau$ is of order $\log \frac{a}{\Lambda}$.
The net number of BPS $2d$ particles in this vacuum is $0$.

As the $z$ parameter approaches the cut, the monodromy parameter increases, up to a point $\alpha \sim 1/2$, where the monodromy felt by the bulk hypermultiplet field is $-1$.
As we pass through the cut and go back to large $z$ on the second branch of vacua, the monodromy parameter goes all the way to $\alpha \sim 1$.
As the same time, the rules of the $2d$ wall-crossing \cite{Gaiotto:2011tf} tell us that the number of $2d$ particles with the same gauge charge as the bulk hypermultiplet
must jump by one unit.

At this point, it is natural to do a large gauge transformation to bring the monodromy parameter back to $\alpha \sim 0$.
A large gauge transformation in the presence of the bulk hypermultiplet matter field is not harmless \cite{Gaiotto:2011tf}: it means that we are trying to replace a setup with one unit of
magnetic flux with a setup with no magnetic flux. In the first setup, the bulk hypermultiplet would have a Landau level localized at the defect, which behaves effectively as a $2d$ particle.
In the latter, it would not. Thus in order to do the large gauge transformation, we must add an extra $2d$ particle. This brings back the net number of BPS $2d$ particles in this vacuum to $0$.
In the original model, before the scaling limit, we could have reached the second vacuum by passing through the other cut in parameter space.
Along such an alternative path, $\alpha$ would have remained continuously small and the $2d$ BPS degeneracy for this charge would not have jumped from zero.
As usual, BPS wall-crossing resolves the possible inconsistencies arising from monodromies in parameter space \cite{Gaiotto:2011tf}.

It is useful to look at a simple toy model to understand how an extra $2d$ particle may help cure the effect of a large gauge transformation on the bulk hypermultiplet field.
As we are not really able to keep track of non-protected physics, it is useful to focus on the modes of the bulk hypermultiplet that are chiral from the point of view of the $2d$ super-algebra.
These are the modes with an holomorphic dependence on the transverse coordinates. We can decompose the hypermultiplet into two complex fields $M$ and $\tilde M$ of opposite gauge charge,
and expand in $2d$ modes $M_k$ and $\tilde M_k$:
\begin{equation}
M = \sum_{k=0}^\infty M_k z^k \qquad \tilde M = \sum_{k=0}^\infty \tilde M_k z^k
~. \end{equation}

If we turn on a background gauge field with small monodromy parameter $\alpha$ and eliminate it with a gauge transformation, the bulk fields acquire twisted boundary conditions:
\begin{equation}
M = \sum_{k=0}^\infty M_k z^{k +\frac{\alpha}{2 \pi}} \qquad \tilde M = \sum_{k=0}^\infty \tilde M_k z^{k -\frac{\alpha}{2 \pi}}
~. \end{equation}
If we keep increasing $\alpha$ all the way to $2 \pi$, we end up with a missing mode in $M$, and an extra mode in $\tilde M$.

On the other hand, suppose we add a $2d$ particle $q$ to the initial system, with superpotential coupling
\begin{equation}
\int d^2 x d^2 \theta q \tilde M|_{z=0}
~. \end{equation}
The F-term equations set
\begin{equation}
\tilde M|_{z=0} =0 \qquad \bar \partial M|_{z=0} = q
~. \end{equation}
Thus the original expansion is modified to
\begin{equation}
M = \frac{q}{z} + \sum_{k=0}^\infty M_k z^k \qquad \tilde M = \sum_{k=1}^\infty \tilde M_k z^k
~. \end{equation}
If we then turn on the monodromy parameter all the way from $0$ to $2\pi$, we go back to the standard situation
\begin{equation}
M = q + \sum_{k=0}^\infty M_k z^{k+1} \qquad \tilde M = \sum_{k=1}^\infty \tilde M_k z^{k-1}
~. \end{equation}

Of course, this is not quite what happens in our model, where the effective monodromy parameter varies continuously and the extra particle is injected as $\alpha$ hits $\pi$.
This is not necessarily a problem. For example, we could renormalize the superpotential coupling to something like
\begin{equation}
\int d^2 x d^2 \theta q  \tilde M_0 ~,
\end{equation}
which would still make sense at generic $\alpha$.
We should also keep in mind that we cannot really trust the effective GW description of the defect in the transition region, where the $2d$ dynamics is clearly intricate, and involves
an interesting spectrum of light $2d$ BPS domain walls. The actual profile of the fields in the region close to the defect  is out of our control.

Putting all pieces together, we arrive to the following collection of statements:
\begin{itemize}
\item Near the monopole point, we have a very weakly-coupled four-dimensional Abelian gauge theory coupled to a bulk hypermultiplet of electric charge $1$ and to some two-dimensional degrees of freedom, which are possibly strongly interacting.
\item The two-dimensional degrees of freedom should include a superpotential coupling to the bulk hypermultiplet involving some electrically charged $2d$ chiral operator $\CO$:
\begin{equation}
\int d^2 \theta d^2 x \CO \tilde M|_{z=0}
~. \end{equation}
\item The two-dimensional degrees of freedom should have two vacua, with appropriate low-energy twisted superpotential and twisted chiral ring.
\item In the absence of the coupling to the bulk hypermultiplet, a $2d$ chiral particle should appear in the second vacuum only, and $\CO$ should
go to the corresponding $2d$ free field $q$ in the IR.
\end{itemize}

\subsection{A special $2d$ theory}
Our next task will be to find a two-dimensional field theory with exactly these properties.
We propose a candidate: a $2d$ $U(1)$ gauged linear sigma model with a charge $1$ field $w_+$ and a charge $-1$ field $w_-$ deformed by a bare twisted superpotential $- \frac{\sigma^2}{4 m}$, the simplest irrelevant
deformation we can add to the local model.

Before the deformation, the theory has a single vacuum. The twisted chiral relation is linear
\begin{equation}
\sigma-\sqrt{u}=  e^{\tilde t} (\sigma-m)
\end{equation}
and gives an effective twisted superpotential, which includes an $-(\sqrt{u}-m) \log (\sqrt{u}-m)$ term,
signalling the presence of a $2d$ BPS particle with the same charge as the light bulk BPS particle.
The $2d$ BPS particle is clearly associated to the $\CO = w_+ w_-$ chiral operator.

After the deformation,  the D-term equations are modified to
\begin{equation}
|w_+|^2 - |w_-|^2 = \tilde t + \frac{\sigma}{2m}
\end{equation}
and thus can be solved in two ways:
\begin{itemize}
\item We can keep $\sigma=0$ and turn on appropriate vevs for $w_\pm$. This is the standard vacuum, with a flat direction which gives the $2d$ chiral field of flavor charge $1$.
\item We can set $\sigma = - 2 m \tilde t$. As $\sigma$ has a vev, the $2d$ chirals become massive, and no $2d$ BPS particle is present.
\end{itemize}

The twisted chiral ring equations for this exotic model
\begin{equation}
\sigma-\sqrt{u}=  e^{\tilde t+ \frac{\sigma}{2m}} (\sigma-m)
\end{equation}
is somewhat unusual, because it has infinitely many solutions, roughly spaced by multiples of $4 \pi i m$. They correspond to vacua with
various amount of $2d$ electric flux. As we scale $m$ to be very large
as before, we can focus on the two vacua near the origin and reproduce the desired rescaled twisted chiral ring relation (\ref{eq:A1s}).

Alternatively, we can reproduce this local model with a small trick: as the features of this theory are not expected to depend
on the specific bulk theory or on the location of the monopole point in the Coulomb branch, we can look at a bulk theory with a logarithmic singularity with a semiclassical description. The $SU(2)$ gauge theory with one flavor does the job nicely.

Recall the twisted chiral ring for the $\mathbb{CP}^1$ surface defect in the $SU(2)$ gauge theory with one flavor:
\begin{equation}
\sigma^2 = e^t + u  + \Lambda^3 (\sigma - m) e^{-t}
~. \end{equation}
The branch locus is at the value of $u$, where the following expression is zero:
\begin{equation}
p(e^t) = e^t + u -m \Lambda^3 e^{-t} + \frac{\Lambda^6}{4} e^{-2 t}
\end{equation}
and the logarithmic singularities occur at values of $u$, where two roots of $p$ collide.

For large $m$, and $u \sim m^2$, we have a logarithmic singularity with a semiclassical interpretation: one of the fundamental bulk hypermultiplets is light.
The branch locus approaches this logarithmic singularity if $t$ approaches the colliding roots of $p$, which are the two roots for which the $e^t$ term in $p$ is negligible.

Let us scale around this point. The bulk mass $m$ can be our large parameter. We can take
\begin{equation}
u = m^2 + m a \qquad e^t = \frac{\Lambda^3}{2 m}(1+ m^{-1/2} z) \qquad \sigma = m + m^{1/2} x
\end{equation}
and derive the twisted chiral ring relation
\begin{equation} \label{eq:A1s}
x^2 + 2 z x = a
~. \end{equation}
This can be matched to the usual local expression by a further shift $x \to x - z$, but we will not do so.

Notice that we are scaling in the region on the second branch of $T(\sigma)$ near the hidden poles at $\sigma =m$. This suggest that we should trade
the $\mathbb{CP}^1$ description of the surface defect for the alternative description for which the second branch of $T(\sigma)$ is semiclassical.
This is the description involving a doublet of $2d$ gauge charge $-1$ together with an extra $2d$ chiral field of $2d$ gauge charge $1$ and $4d$ flavor charge $1$.
This two-dimensional field content should give a rather faithful description of the physics in the region of interest. For example,
the correction to the twisted chiral ring due to the bulk dynamics is captured by the $e^t$ term in the twisted chiral ring, which is completely
suppressed in the scaling region.

We are precisely scaling our parameters in such a way that both the extra $2d$ chiral and one of the two chirals in the doublet are relatively light,
while the second chiral in the doublet has a mass of order $2m$. We can
integrate out the massive $2d$ chiral field and expand its contribution to the effective twisted superpotential in inverse powers of the mass. The first non-trivial term after the
FI parameter renormalization is the desired bare twisted superpotential $- \frac{\sigma^2}{4 m}$.

\section{Symmetries and the displacement multiplet}
The BPS surface defects we study in this paper break several of the symmetries of the bulk theory.
Clearly, they break translations in the transverse $x^2$ and $x^3$ directions. They break the $SU(2)_R$ symmetry of the bulk theory down to an $U(1)_R$ Cartan subgroup.\footnote{This should not to be confused with the $U(1)_r$ bulk R-symmetry of a conformal bulk theory, which will be independently preserved by conformally invariant defects.} They also break half of the supercharges. The preserved supercharges can be identified with the subset of bulk charges with the same quantum numbers under $U(1)_R$ and the rotations
$U(1)_{23}$ around the defect.

In particular, the $U(1)_d$ symmetry generated by the difference between the generators of $U(1)_{23}$ and $U(1)_R$
behaves in many ways as a two-dimensional non-R flavor symmetry. This symmetry is used, for example,
to introduce an extra grading in BPS state counting \cite{Gaiotto:2010be,Gaiotto:2011tf}.
Notice, within a given $4d$ $\CN=2$ multiplet fields with the same $U(1)_d$ charge
compose a supermultiplet for the $(2,2)$ sub-algebra preserved by the defect,
thereby providing a decomposition of $\CN=2$ multiplets into $(2,2)$ supermultiplets.

In the rest of the paper, we will need several details of how bulk fields transform
under the symmetries preserved or broken by the defects. It is convenient to use superfields
for some calculations. Vectormultiplets can be described readily by a naive $\CN=2$ superspace,
and then decomposed into $(2,2)$ superfields. Hypermultiplets cannot be described as readily
in $\CN=2$ superspace, but can be readily decomposed into $(2,2)$ superfields.

A vectormultiplet can be described as a chiral $\CN=2$ superfield,
\begin{equation}
\Phi = \phi + \theta^\alpha_i \lambda_\alpha^i + \cdots
~. \end{equation}
Latin indices are doublet indices for $SU(2)_R$. We denote as $D^i_\alpha$ the usual superspace derivatives.\footnote{Often, a Bianchi identity is also imposed, restricting $D^{(i} D^{j)} \Phi$ to be real. }
We label spinors by their quantum numbers under the $2d$ boost along the defect $Spin(1,1)_{01}$ and the rotation $U(1)_{23}$ in the plane orthogonal to the defect as follows. The two components of a $4d$ left-handed spinor are denoted as $+,+$ and $-,-$. And the two components of a $4d$ right-handed spinor are denoted as $+,-$ and $-,+$.
In these conventions, a half-BPS defect preserves supercharges $Q_{-,+}^+$ and $Q_{+,+}^+$ and their conjugates, $Q_{-,-}^-$ and $Q_{+,-}^-$
(which in the literature on $2d$ $\CN=(2,2)$ theories are often denoted $\overline{Q}_-$ and $\overline{Q}_+$ respectively, {\it cf} Table \ref{tab:super}).

We can obtain $(2,2)$ superfields simply by taking chiral superspace derivatives $D^+_{-,-}$ and $D^-_{+,+}$ in the broken directions and then restricting the
corresponding chiral superspace coordinates $\theta_+^{-,-}$ and $\theta_-^{+,+}$ to zero. The resulting $(2,2)$ multiplets are still annihilated by the
anti-chiral unbroken supercharges $Q^+_{-,+}$ and $Q^-_{+,-}$, i.e.\  are all $(2,2)$ twisted chiral multiplets.
We find the familiar $U(1)_d$-neutral twisted chiral multiplet we encountered in twisted F-terms
\begin{equation}
\Phi |_{\theta_+^{-,-} = \theta_-^{+,+}=0} = \phi +  \theta^{+,+}_+ \lambda_{+,+}^+ + \theta^{-,-}_- \lambda_{-,-}^-  +\cdots \end{equation}
two fermionic twisted chiral multiplets of $U(1)_d$ charge $\pm 1$
\begin{equation}
\qquad D^+_{-,-}\Phi |_{\theta_+^{-,-} = \theta_-^{+,+}=0}= \lambda^+_{-,-} + \cdots
\qquad D^-_{+,+}\Phi |_{\theta_+^{-,-} = \theta_-^{+,+}=0}= \lambda^-_{+,+} + \cdots
\end{equation}
and one more bosonic twisted chiral multiplet $D^-_{+,+} D^+_{-,-}\Phi |_{\theta_+^{-,-}= \theta_-^{+,+}=0}$.

As for hypermultiplets, we can denote the scalar fields as $M_i^A$, with an R-symmetry index and a symplectic flavor index $A$ and appropriate reality condition
\begin{equation} (M^\dagger)_A^i = \epsilon^{ij} \omega_{AB} M^B_i ~. \end{equation}
The SUSY transformations read schematically
\begin{equation}
Q^i_\alpha M_j^A = \delta^i_j \rho^A_\alpha \qquad \bar Q^i_{\dot \alpha} M_j^A = \delta^i_j \bar \rho^A_{\dot \alpha}
\end{equation}
and thus the unbroken supercharges $Q^+_{\pm,+}$ annihilate $M_-^A$, which forms a $(2,2)$ chiral multiplet together with $\rho^A_{\pm,-}$.
The scalar $M_+^A$ belongs to the complex conjugate $(2,2)$ anti-chiral multiplet.

\begin{table}
\begin{centering}
\begin{tabular}{|c|c|c|c|}
\hline
~4d $\CN=2$ theory~ & ~$(2,2)$ defect~ & ~4d $\CN=1$ deformation~ & ~$(2,0)$ defect~ \tabularnewline
\hline
\hline
$Q_{-,-}^-$ & $\overline{Q}_-$ & $Q_{\alpha = 1}$ & $\overline{Q}_-$ \tabularnewline
\hline
$Q_{+,+}^-$ & broken & $Q_{\alpha = 2}$ & broken \tabularnewline
\hline
$Q_{-,+}^+$ & $Q_-$ & $\overline{Q}_{\dot \alpha = \dot 1}$ & $Q_-$ \tabularnewline
\hline
$Q_{+,-}^+$ & broken & $\overline{Q}_{\dot \alpha = \dot 2}$ & broken \tabularnewline
\hline
$Q_{+,+}^+$ & $Q_+$ & broken & broken \tabularnewline
\hline
$Q_{-,-}^+$ & broken & broken & broken \tabularnewline
\hline
$Q_{+,-}^-$ & $\overline{Q}_+$ & broken & broken \tabularnewline
\hline
$Q_{-,+}^-$ & broken & broken & broken \tabularnewline
\hline
\end{tabular}
\par\end{centering}
\caption{\label{tab:super}Supercharges of the $4d$ $\CN=2$ theory
preserved by the defect and $\CN=1$ deformation, separately and together.
We compare our notations to those used in the literature ({\it e.g.} \cite{Witten:1993yc}
in the second and fourth column, and \cite{Wess:1992cp} in the third column.)}
\end{table}

\subsection{The displacement supermultiplet}
The Ward identities for symmetries broken by a defect are not lost. Rather, they
hold away from the defect, and receive corrections that are localized on the defect itself.
The typical example is a translation in the directions transverse to the defect:
\begin{equation}
\partial_\mu T^{\mu 2} =  \delta(x^2)\delta(x^3) \,   {\bf d}^2 \qquad , \qquad\partial_\mu T^{\mu 3} =  \delta(x^2)\delta(x^3) \, {\bf d}^3
~. \end{equation}
This Ward identity defines the displacement operators ${\bf d}^{2,3}$ on the defect. Alternatively, the displacement operators can be thought as operators that can be added to a defect Lagrangian in order to shift the location of the defect.

In a similar fashion the Ward identity for the off-diagonal generators of $SU(2)_R$
\begin{equation}
\partial_\mu (J^+)^{\mu} =  \delta(x^2)\delta(x^3) \, r^+ \qquad  , \qquad \partial_\mu (J^-)^{\mu} =  \delta(x^2)\delta(x^3) \, r^-
\end{equation}
defines operators $r^\pm$ on the defect with $U(1)_R$ charge $\pm 1$, which can be added to the defect Lagrangian in order to rotate
the choice of unbroken subgroup of $SU(2)_R$. In particular, they are exactly marginal operators!

Finally, the four broken supersymmetries are associated with four fermionic operators on the defect.
The four operators on the defect corresponding to the broken supercharges have
opposite $\pm 1/2$ charges under $U(1)_R$ and $U(1)_{23}$. Notice that all the broken currents and associated operators on the defect have $U(1)_d$ quantum number $\pm1$, and belong to a complex conjugate pair of supermultiplets of the $(2,2)$ superalgebra preserved by the defect.
Thus we will refer to all these operators on the defect as ``displacement operators''.

We can be more precise about the supersymmetry transformation rules for the displacement operators.  The two unbroken supercharges with positive $U(1)_R$ charge
annihilate $r^+$.
The other two supercharges map $r^+$ to two fermionic operators of $U(1)_R$ charge $1/2$ and $U(1)_{23}$ charge $-1/2$. Acting again with these supercharges, we should arrive to ${\bf d}^2 - i {\bf d}^3$, the complex combination of displacement operators of $U(1)_{23}$ charge $-1$.
Thus the displacement operators form a chiral supermultiplet $R^+$ with charge $-1$ under $U(1)_d$.
All the fields in our $(2,2)$ theory are neutral under $U(1)_d$.  Therefore, the operators in $R^+$ must include additional bulk fields that are charged under this symmetry.

For simplicity, consider a surface defect described by a 2d Lagrangian such that the bulk degrees of freedom only enter 2d twisted F-terms.
This is the case for the IR GW description of surface defects, or for the intermediate effective Lagrangian we obtain after integrating away the 2d chiral fields.
Working in superspace,
the Lagrangian coupling associated to some twisted superpotential $\CW$
is \begin{equation}
D^+_{+,+} D^-_{-,-} \CW
~. \end{equation}
Acting with the $SU(2)_R$ raising operator that increases the $U(1)_R$ charge by $+1$ we find
\begin{equation}
r^+ = D^+_{+,+} D^+_{-,-} \CW
~, \end{equation}
where the $D^+_{-,-}$ operator acts trivially on the $2d$ degrees of freedom and must act on the bulk fields.
Thus $r^+$ is really a $(2,2)$ descendant of a fermionic twisted chiral operator $D^+_{-,-} \CW$.

Consider first the twisted effective superpotential that gives the IR description of a surface defect.
Starting from $\CW$, which has zero $U(1)_R$ charge and integrating over the twisted chiral superspace,
we get two terms in the Lagrangian, both of zero $U(1)_R$ charge: the actual GW couplings
\begin{equation}
(F_{01} + iF_{23}) \, \partial \CW
\end{equation}
and the fermion bilinear
\begin{equation}
\lambda^+_{+,+}\lambda^-_{-,-} \, \partial^2 \CW ~,
\end{equation}
which involves the four-dimensional gaugino components that have the same charge under $U(1)_R$, $U(1)_{23}$ and boost along the defect $Spin(1,1)_{01}$.
While the former term is $SU(2)_R$ invariant, the latter is not. It is in the same $SU(2)_R$ multiplet as
\begin{equation}
r^+_{\CW} = \lambda^+_{+,+}\lambda^+_{-,-}\partial^2 \CW = \frac{1}{2} \lambda^+_\alpha \lambda^{+,\alpha} \partial^2 \CW
~. \end{equation}
Here we used the fact that the $(+,+)$ and $(-,-)$ components of the Lorentz spinor index are the $4d$ chiral components, and thus the product can be written as a Lorentz invariant contraction.
This is an interesting operator.  First, it is the restriction to the defect of a scalar bulk operator. From the point of view of the bulk theory, this operator
is simply a descendant of the Coulomb branch operator $\CW$. If we focus on the $\CN=1$ subalgebra of the bulk superalgebra generated by the $Q^-_\alpha$ generators and their complex conjugates,
we find that $r^+_{\CW}$ is the restriction to the defect of an $\CN=1$ chiral field. This is interesting because it will allow us to make contact with the results of the anomaly equations analysis of \cite{Cachazo:2002ry,Seiberg:2002jq,Cachazo:2003yc}.

With that purpose in mind, it is useful to specialize to the basic surface defect defined by a chiral fundamental coupled to an $SU(N)$ gauge theory.
If we do the calculation after integrating out the doublet, so that the defect Lagrangian is controlled by the effective twisted superpotential
\begin{equation}
2\pi i \CW =- \mathrm{Tr} \left(m+ \Phi \right) \log \left(m + \Phi \right)/e
\end{equation}
we find
\begin{equation}
r^+_{\CW} = \tilde R(m) \equiv -\frac{1}{4\pi i}\mathrm{Tr} \frac{\lambda^+_\alpha (\lambda^+)^\alpha}{m + \Phi} - \frac{1}{2\pi i}\mathrm{Tr} D^+ \log \left(m + \Phi \right) ~,
\end{equation}
where $D^+$ is a bulk auxiliary field.
For a pure $SU(N)$ gauge theory, we can drop the auxiliary field and identify $r^+ (m)$ with the generating function $R(m)$
of the $\CN=1$ chiral operators $\mathrm{Tr} \CW_\alpha \CW^\alpha \Phi^k$, which plays an important role in the
analysis of the anomaly equations  \cite{Cachazo:2002ry,Seiberg:2002jq,Cachazo:2003yc}. For a theory with bulk matter, the second term can be recast in a useful form by replacing the
auxiliary field with the appropriate bilinear of bulk hypermultiplets, roughly of the form $\tilde M \log \left(m + \Phi \right) M$,
which is also used in the analysis of the anomaly equations  \cite{Cachazo:2002ry,Seiberg:2002jq,Cachazo:2003yc}. We denote the full expression as $\tilde R(m)$ to highlight the close relationship to $R(m)$.

It is worth mentioning that displacement operators play an interesting role in the identification of surface defects in some theory with the low energy limit
of some dynamical vortices in a larger UV completion of the theory. A vortex will always support some zero-modes corresponding to the broken translation symmetry,
which should remain as massless excitations in the IR. For a BPS vortex these massless excitations are organized into a chiral multiplet, carrying $U(1)_d$ charge $1$.
It should be clear that at the linearized order, this multiplet $\delta x$ must interact with the rest of the theory through a coupling to the displacement operator, i.e.\
a superpotential coupling of the rough form
\begin{equation}
\int d^2\theta R^+ \delta x
~. \end{equation}
We expect this coupling to be marginally irrelevant, so that the vortex theory in the IR can be described by some surface defect very weakly coupled to the free chiral
$\delta x$. Such a free chiral is invisible in twisted chiral computations, but is visible, say, in index computations \cite{Gaiotto:2012xa,Gadde:2013dda}.
Even the ``basic surface defect'' discussed in the previous paragraph has two variants defined by coupling to the $\mathbb{CP}^{N-1}$ sigma-model in two dimensions, with or without the
extra chiral multiplet, which is neutral under the $4d$ gauge group and is coupled to the displacement operators. In the context of vortex strings,
these two possibilities were discussed in \cite{Shifman:2005st} and \cite{Edalati:2007vk}, respectively.

\section{$\CN=1$ symmetry breaking}
Until now, we have found that several natural operators on a basic surface defect have a close relationship to the
bulk $\CN=1$ chiral operators that participate in the anomaly equations for $\CN=2$ theory broken to $\CN=1$
by extra superpotential terms. It is natural to ask if our results may allow us to understand the behavior of surface defects
under such a bulk deformation of the theory. We will focus on bulk superpotential deformations involving Coulomb branch
operators of the $\CN=2$ theory.

\subsection{Warmup: a non-renormalization theorem in the bulk}
Consider the following {\it pre-potential} deformation
\begin{equation} \label{eq:N1def}
\int d^4 \theta E W[\Phi]
\end{equation}
for some gauge-invariant polynomial $W[\Phi]$ of the vectormultiplet scalars.
We can promote the coefficient $E$ to an $\CN=2$ vector superfield and give a vev to a very specific scalar component of the $E$ superfield:
\begin{equation}
E = \cdots + \theta_+^{+,+}\theta_+^{-,-} \epsilon + \cdots
~. \end{equation}
This converts the pre-potential deformation into an $\CN=1$ superpotential deformation
for the $\CN=1$ subalgebra defined by the $Q^-_\alpha$ and their complex conjugates (see Table \ref{tab:super} for the list of unbroken generators):
\begin{equation}
\int d \theta_-^{+,+} d \theta_-^{-,-} \epsilon W[\Phi]
~. \end{equation}
Generically, such a deformation lifts the moduli space of vacua except for a finite set of points determined by the parameters in $W[\Phi]$.
Our goal, then, is to analyze how the vevs of $\CN=1$ primaries that come from $\CN=2$ primaries (and their descendants)
vary with the parameters of $W[\Phi]$ in a given vacuum.

Intuitively, we expect that at the leading order the effect of this $\CN=1$ deformation will be related to the effect of the original pre-potential deformation.
For example, if the vev of a Coulomb branch operator $\CO$ would shift by $E \delta \CO$ under the pre-potential deformation,
we should be able to promote that statement to a superfield statement, and read off the variation under the superpotential deformation
of the vev of the descendant $D_{+,+}^+ D_{-,-}^+ \CO$. As the initial vev is zero, the final vev under the $\CN=1$ deformation should be
\begin{equation} \label{eq:bulkdef}
D_{+,+}^+ D_{-,-}^+ \CO = \epsilon \delta \CO
~. \end{equation}

We can extend this result to be exact and make it into a non-renormalization theorem.
The $\CN=1$ chiral superfields included in an $\CN=2$ vector multiplet
are $\Phi$, $D_\alpha^+ \Phi$, and $D_{\alpha}^+ D^{+ \alpha} \Phi$. All except $\Phi$ have positive $U(1)_R$ charge.
The hypermultiplets contain chiral superfields $M_-^A$, which also have positive $U(1)_R$ charge (remember we lower $SU(2)_R$ indices with an $\epsilon$ tensor).
The vevs of chiral multiplets should be holomorphic in $\epsilon$, which has $U(1)_R$ charge $1$.
Thus chiral operators which carry negative or zero $U(1)_R$ charge (in particular, the original twisted chiral multiplets)
should acquire no vev, while operators of positive $U(1)_R$ charge should acquire vevs proportional to appropriate powers of $\epsilon$.

In particular, the right hand side of the chiral equation (\ref{eq:bulkdef}) must be linear in $\epsilon$ (which also has $U(1)_R$ charge $1$)\footnote{Our conventions are such that the supercharges and the $\theta$s have $U(1)_R$ charges $\pm \frac{1}{ 2}$.  This is natural in $\CN=2$, but is not the standard convention in $\CN=1$ supersymmetry, where their charges are $\pm 1$.}
and uncorrected from the first order answer.
This expectation is realized in the calculations based on anomaly equation: the vevs of operators such as $\Tr \lambda_\alpha \lambda^\alpha \Phi^k$
are linear under a rescaling of the superpotential deformation. It would be interesting to compare directly
the vevs under $\CN=1$ deformation with the changes of the vevs under pre-potential deformations of the $\CN=2$ theory.

\subsection{A non-renormalization theorem on the defect}
We can apply the same ideas to the response of half-BPS surface defects to
the $\CN=1$ bulk deformation. In general, a pre-potential deformation (\ref{eq:N1def}) changes the vevs of $(2,2)$ twisted chiral
multiplets. The superfield $E$ contains several $(2,2)$ twisted chiral superfields, and each can contribute to the
change in the vevs. The $\epsilon$ deformation term is contained in the fermionic superfield
\begin{equation}
\eta = D^+_{-,-}E |_{\theta_+^{-,-} = \theta_-^{+,+}=0} =  \cdots + \theta_+^{+,+} \epsilon + \cdots
~. \end{equation}
Notice that $\epsilon$ has $U(1)_d$ charge $-1$.  Thus the bulk superpotential deformation is akin to a deformation by a level one descendant of a twisted chiral field.

The $\CN=1$ bulk deformation, combined with the symmetry breaking due to the surface defect,
reduces the supersymmetry of the system to $(2,0)$ in two dimensions,
with left-moving\footnote{With our choice of conventions, summarized in Table \ref{tab:super}, the unbroken
supersymmetry resides in the left-moving sector, therefore leading to a $(2,0)$ theory in two dimensions.
This choice is opposite to a more traditional, right-moving supersymmetry used in the literature on $(0,2)$ models and heterotic strings.}
supercharge $Q_{-,+}^+$ and its conjugate $Q_{-,-}^-$ (which in the literature are often denoted $Q$ and $\bar Q$, respectively).
In addition, the resulting two-dimensional $(2,0)$ theory enjoys an R-symmetry $U(1)_R$ and an extra flavor symmetry $U(1)_d$,
both of which are preserved by the $\CN=1$ bulk deformation and the surface defect,
and which will help us in what follows to gain control over our $(2,0)$ theory on the defect.
This is somewhat similar to the previous study of heterotic chiral rings \cite{Katz:2004nn,Adams:2005tc},
where the presence of extra global symmetries (that can play the role of R-symmetries) and the $(2,2)$ locus
were important ingredients for the existence of finite-dimensional $(2,0)$ chiral rings.
While below we follow a different approach, which is tailored for the study of $(2,0)$ defects in 4d $\CN=1$ gauge theories
and does not require any assumptions of conformal invariance, our conclusions are compatible with those of \cite{Adams:2005tc}.

Before we take the full advantage of the extra symmetry $U(1)_d$, though, let us derive some general properties of $(2,0)$ theories.
Their chiral operators $\CO$ are in $Q_-$ cohomology (remember, in our applications we identify $Q_- \equiv Q_{-,+}^+$) \footnote{In order for the definition of chiral operator in $(2,0)$ to agree with twisted chiral in $(2,2)$ and chiral in $\CN=1$ we use $Q_-$ here instead of the more conventional $\bar Q_-$.}
\begin{equation}
[Q_{-,+}^+ \; ,\;  \CO] = 0 \qquad , \qquad \CO \not= \{ Q_{-,+}^+ \; , \; \cdot \}
~. \end{equation}
(The commutators and anticommutators are exchanged if $\CO$ is fermionic.)
Consider a correlation function of such chiral operators
\begin{equation}
F(z_i, \tilde z_i) \; = \; \langle \, \CO_1(z_1,\tilde z_1) \, \CO_2(z_2,\tilde z_2) \, \ldots \, \rangle ~,
\end{equation}
where $\tilde z_i$ is the complex conjugate of $z_i$ in Euclidean signature and it is an independent real coordinate in Lorentzian signature.
As is standard in situations with more supersymmetries,
\begin{align}
\partial_{\tilde z_1} F(z_i, \tilde z_i)  &= \langle [\{ Q_{-,-}^- \; ,\; Q_{-,+}^+ \}, \; \CO_1(z_1,\tilde z_1)] \; \CO_2(z_2,\tilde z_2) \, \ldots \, \rangle \cr
& = \langle \{ Q_{-,+}^+ \; , \; [ Q_{-,-}^- \; , \; \CO_1(z_1,\tilde z_1)] \} \; \CO_2(z_2,\tilde z_2) \, \ldots \, \rangle \cr
& = \langle \{ Q_{-,+}^+ \; , \; [ Q_{-,-}^- , \CO_1(z_1,\tilde z_1)] \; \CO_2(z_2,\tilde z_2) \, \ldots \, \}\rangle =0 ~.
\end{align}
Hence, $F(z_i,\tilde z_i)$ is independent of $\tilde z_i$; i.e.\ it is a meromorphic function of $z_i$.

Note that unlike the situation with both left and right moving supersymmetries (or the analogous situation in higher dimensions), we cannot prove that the correlation functions of chiral operators are constant.\footnote{In fact it is easy to construct $(2,0)$ theories for which this is not true. For example, consider a $(2,2)$ theory and tensor it with a purely holomorphic conformal field theory, e.g.\ the $E_8$ theory.  Then, there are infinitely many chiral operators for which the correlation function $F$ is not a constant.}

As in theories with more supercharges, it is easy to prove that the correlation function $F$
is holomorphic in coupling constants in the superpotential.  A general superpotential coupling is of the form
\begin{equation}
\int d \theta \epsilon  \Psi  + c.c. ~,
\end{equation}
where $\Psi$ is a fermionic chiral superfield, typically a product of a fermionic chiral superfield times a function of bosonic chiral superfields.  Then,
\begin{align}
\partial_{\bar \epsilon} F(z_i) & = \int d^2 z\langle \{ Q_{-,+}^+ \; , \; \Psi(z ,\tilde z)\} \; \CO_1(z_1,\tilde z_1)] \; \CO_2(z_2,\tilde z_2) \, \ldots \, \rangle \cr
& = \int d^2 z \langle \{ Q_{-,+}^+ \; , \; \Psi(z,\tilde z) \, \CO_1(z_1,\tilde z_1) \, \CO_2(z_2,\tilde z_2)  \, \ldots \, \}\rangle =0 ~.
\end{align}
A similar argument shows that $F(z_i)$ cannot depend on any non-chiral parameter -- a parameter that deforms terms
in the Lagrangian of the form $\int d^2 \theta \chi$ with $\chi$ a generic superfield.
We conclude that $F$ can depend only on chiral parameters $\epsilon$.

The $(2,0)$ theories we find in our setup are not the generic ones.
They are continuous deformations of $(2,2)$ theories on the defect. Therefore, they enjoy some additional properties.
Their chiral operators are of three kinds.  First, $(2,2)$ chiral operators $\Phi$ are $(2,0)$ chiral operators.
Second, $(2,2)$ twisted chiral operators $\Sigma$ are $(2,0)$ chiral operators.
And third, there are many other $(2,0)$ chiral operators some of which are $(2,2)$ descendants of $\Phi$ and $\Sigma$.
Namely, each $(2,2)$ chiral multiplet $\Phi$ decomposes into two $(2,0)$ chiral multiplets, $\Phi$ and $D_{+,+}^- \Phi$, of opposite statistics.
A $(2,2)$ twisted chiral multiplet $\Sigma$ also decomposes into two $(2,0)$ chiral multiplets, $\Sigma$ and $D_{+,+}^+ \Sigma$ in a similar way.

At the $(2,2)$ point, which we will denote by $\epsilon=0$, the correlation functions
\begin{align}
F_\Phi(z_i, \tilde z_i) & = \langle \, \Phi_1(z_1,\tilde z_1) \, \Phi_2(z_2,\tilde z_2)  \, \ldots \, \rangle |_{\epsilon=0} \cr
F_{\Sigma}(z_i, \tilde z_i) & = \langle \, \Sigma_1(z_1,\tilde z_1) \, \Sigma_2(z_2,\tilde z_2)  \, \ldots \, \rangle |_{\epsilon=0}
\end{align}
are independent of the positions $z_i,\tilde z_i$.
More generally, the $(2,2)$ chiral operators $\Phi$ and the $(2,2)$ twisted chiral operators $\Sigma$ form two separate rings.
Note that other correlation functions of $(2,0)$ chiral fields that are not of the form $F_\Phi$ or $F_{\Sigma} $ are not necessarily constants.

Next we would like to understand what happens to these correlation functions when $\epsilon$ is nonzero. The key point is that the parameter $\epsilon$ carries nontrivial charge under $U(1)_d$, and all other chiral operators or parameters in the theory carry a $U(1)_d$ charge of the same sign as $\epsilon$.  Therefore, since the correlation functions of all $(2,0) $ chiral operators are holomorphic in $\epsilon$, the correlation function of chiral operators with zero $U(1)_d$ charge must be independent of $\epsilon$.  Hence, they are given by their values for $\epsilon=0$; i.e.\ they are precisely as in the un-deformed $(2,2)$ theory.  In particular, the twisted chiral ring of the $(2,2)$ theory is not modified by nonzero $\epsilon$, and hence the space of vacua and the vevs of these operators are exactly as in the $(2,2)$ theory.\footnote{Here it is crucial that our defect is not a dynamical excitation of the system.  Otherwise, there would have been a dynamical displacement mode, whose $U(1)_d$ charge could be opposite to that of $\epsilon$, and our whole reasoning would have failed.}

It follows that the $(2,0)$ fermionic superpotential of the deformed theory is unaffected by the bulk $\CN=1$ deformation
and can be determined at the $(2,2)$ point. This fermionic superpotential has a general structure
\begin{equation}
\Lambda^a J_a [\Sigma^a] ~,
\end{equation}
where $\Lambda^a$ are the $(2,0)$ fermionic chiral multiplets and $J_a [\Sigma^a]$ are holomorphic functions
of the $(2,0)$  bosonic chiral multiplets $\Sigma^a$. On the $(2,2)$ locus, each pair of $\Sigma^a$ and $\Lambda^a$
combine into a complete $(2,2)$ twisted chiral multiplet, and $J_a [\Sigma^a] = \partial_a \CW[\Sigma^a]$.
The above arguments show that this fermionic superpotential is uncorrected by the $(2,0)$ deformation
and the vacua of the theory are still described by the extremum equations
\begin{equation}
J_a = \partial_a \CW[\Sigma^a] =0
\end{equation}
even away from the $(2,2)$ locus.

This line of reasoning is somewhat new. We are used to deriving exact results in supersymmetric theories using symmetries that are not there; i.e.\  explicitly broken symmetries. We are also used to deriving approximate results when supersymmetry is broken. Here we find exact results, which follow from an explicitly broken global symmetry and explicitly broken supersymmetry. The point is that the breaking parameter $\epsilon$ is the bottom component of a $(2,0)$ chiral superfield and it is charged under a non-R-symmetry and no other chiral superfield is charged under it.

Next, we will verify these ideas by looking at brane realizations of the $\CN=1$ deformations and surface defects.

\subsection{Brane realization of $\CN=1$ deformations}

The standard setup to engineer $\CN=2$ gauge theories involves D4 branes stretched between parallel NS5 branes \cite{Witten:1997sc}.
The NS5 branes extend along the space-time directions ant two extra directions, say $x^4$ and $x^5$. They are separated along the $x^6$ direction.
A well known way to break the supersymmetry of the system further to $\CN=1$ is to
deform the NS5 brane configuration further \cite{Witten:1997ep}, to wrap holomorphic curves in the two-dimensional complex space parameterized by
\begin{equation}
v = x^4 + i x^5 \,, \qquad
w = x^8 + i x^9 \,.
\end{equation}

We will at first consider an example with two NS5 branes only, defining a pure $SU(N)$ gauge theory,
and then extend it to many NS5 branes.
\subsubsection{Pure $SU(N)$}

\begin{figure}[t] \centering
\includegraphics[width=2.5in]{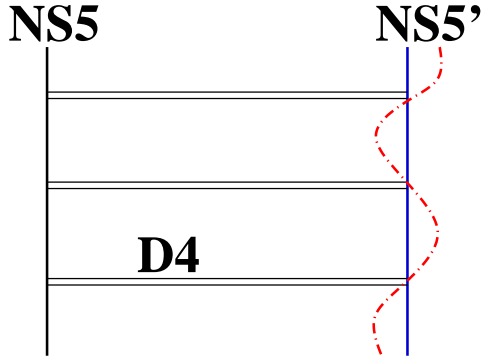}
\caption{\label{N1branes}
The brane construction of $\CN=2$ super Yang-Mills theory
softly deformed to $\CN=1$ by the superpotential $\Tr \, W (\Phi)$.
On the right, the blue line represents the original NS5-brane (supported at $w=0$)
and the squiggly red line represents the deformed NS5-brane (supported on the curve $w = W'(v)$).}
\end{figure}

We can consider a deformation of the form $w = f(v)$ for the rightmost NS5 brane \cite{Witten:1997ep}, as in figure \ref{N1branes}.
Classically, the D4 brane segments will adapt super-symmetrically to the NS5 branes deformation
by sitting at a value of $v$ where $f(v) = 0$, so that they can have a definite position in the $w$ direction.

We can interpret this classical behavior naturally in terms of a superpotential deformation
\begin{equation}
\\Tr W[\Phi] ~,
\end{equation}
where $\Phi$ is the Coulomb branch scalar which describes the motion in the $v$ direction of the
D4 branes. We simply identify
\begin{equation}
f(v) = W'(v)
\end{equation}

After a lift to M-theory, the branes will join into a single M5 brane wrapping an holomorphic curve in the $v,w,t \equiv x^6 + i x^{10}$ directions.
The boundary conditions in the $v,t$ directions are not affected by the deformation, and thus the curve can still be projected to the
$\CN=2$ curve in the $v,t$ coordinates. The only difference is that the curve will have some extra deformation in the $w$ direction,
which can be described by some holomorphic function on the $\CN=2$ curve, constrained by the boundary conditions
\begin{equation}
w \sim 0 \qquad \qquad w \sim f(v)
\end{equation}
whenever $v$ becomes large along the direction corresponding to the first or second NS5 branes, respectively.

Before getting to explicit formulae, we can add to the system a D2/M2 brane, representing a simple surface defect
in the four-dimensional field theory. The D2 brane is essentially semi-infinite, with world-volume along the directions $x^0$, $x^1$, and $x^7 \geq 0$.

The position of the M2 brane in the $v$ direction, which corresponds to the vev of the basic twisted chiral operator $\sigma$,
is unaffected by the $\CN=1$ deformation. The only difference is that the $M2$ brane has to move in the $w$ direction
to remain attached to the deformed M5 brane. What is the field-theory meaning of this movement?
A quick look at the geometry of the system should convince the reader that the displacement of
the D2 brane in the $w$ direction must coincide with the vev of the $r^+$ displacement operator!
Essentially, this follows from the fact that $SU(2)_R$ rotates the $789$ directions.
Thus the action density of the surface defect, i.e.\  the regularized length of the D2 brane in the $7$ direction, changes linearly in the $w$ displacement under $SU(2)_R$ transformations.
We will give a different perspective on this in a later section \ref{sec:twozero}.

Now we are ready to put all pieces together. We know that the surface defect can be described in the neighborhood
of the first NS5 brane in terms of a set of
fundamental chirals coupled to the gauge group. Integrating out the chiral multiplets, and
evaluating the displacement operator we obtain
\begin{equation}
r^+(m) = R(m)
\end{equation}
in terms of the vevs of the $\CN=1$ chiral operators.
Thus we expect the identification
\begin{equation}
v = m \qquad w(v) = r^+(v) = R(v)
\end{equation}

The gauge theory calculations based on the anomaly equations give
\begin{equation}
2R(x) = W'(x) - \sqrt{(W'(x))^2 - g(x)}
\end{equation}
with $g(x)$ a polynomial of degree smaller than $W'$.
The choice of $g(x)$ is constrained so that $R(x)$ has the same branch points as $T(x)$.
This determines the vacua of the system.

For sufficiently large degree of $W$, some vacua are at general locations of the Coulomb branch, and
\begin{equation}
(W'(x))^2 - g(x) = H_1(x)^2 \left( P_N^2 - 4 \Lambda^{2N} \right)
~. \end{equation}
Other vacua appear at monopole points in the Coulomb branch, where we can have
\begin{equation}
(W'(x))^2 - g(x) = H_1(x)^2 G(x) \qquad P_N^2 - 4 \Lambda^{2N}  = H_2(x)^2 G(x)
~. \end{equation}

Clearly, $R(x)$ is a single-valued function
\begin{equation}
2R(v) = W'(v) - H_1(v) \left(e^t - \Lambda^{2N} e^{-t} \right) ~,
\end{equation}
on the M-theory curve, which goes to zero far along the first NS5 brane, and to $W'(v)$ along
the second brane.
It can thus be identified successfully with the displacement of the brane system in the $w$ direction!

\subsubsection{Quiver gauge theories}
Next, we can consider a linear quiver gauge theory, engineered by D4 branes suspended between a sequence
of NS5 branes. We can consider a deformation of the form $w = f_a(v)$ for the $a$-th NS5 brane.
Classically, the D4 brane segments have two options in adapting super-symmetrically to the NS5 branes deformation.
The first possibility is for a brane segment to sit at a value of $v$ where $f_a(v) = f_{a+1}(v)$, so that they can have a definite position in the $w$ direction.
The second is for two or more D4 brane segments to recombine, and separate from the intermediate NS5 branes. For example, two consecutive segments
could recombine and sit at $f_{a-1}(v) = f_{a+1}(v)$, far from the $a$-th NS5 brane.

We can interpret this classical behavior naturally in terms of a superpotential deformation
\begin{equation}
\sum_a \Tr W_a[\Phi_a] ~,
\end{equation}
where $\Phi_a$ is the Coulomb branch scalar which describes the motion in the $v$ direction of the
D4 branes in the $a$-th interval. Indeed, if we identify
\begin{equation}
f_{a+1}(v) - f_{a}(v) = W_{a}'(v)
\end{equation}
the condition $f_a(v) = f_{a+1}(v)$ for a D4 brane correspond to the requirement of eigenvalues of $\Phi_a$ to sit at critical points of $W_a$.
The brane recombination corresponds to vacua where the bifundamental hypermultiplets in the $\CN=2$ quiver gauge theory get a vev:
for example, we can extremize the superpotential terms
\begin{equation}
\Tr W_{a-1}[\Phi_{a-1}] + M_a \Phi_{a-1} \tilde M_a - \tilde M_a \Phi_a M_a + \Tr W_{a}[\Phi_{a}]
\end{equation}
if we set an eigenvalue of $\Phi_a$ equal to an eigenvalue of $\Phi_{a+1}$, so that $M_a$ and $\tilde M_a$ can get an appropriate vev.
We can eliminate $M_a$ and $\tilde M_a$ from the extremum equation to get that the eigenvalues must sit at a critical point of $W_{a-1} + W_a$, i.e.\  at a zero of
$f_{a-1}(v) - f_{a+1}(v)$.

After a lift to M-theory, the deformation in the $w$ direction,
will be described by some holomorphic function on the $\CN=2$ curve, constrained by the boundary conditions
\begin{equation}
w \sim f_a(v)
\end{equation}
whenever $v$ becomes large along the direction corresponding to the $a$-th NS5 brane.

We know that the surface defect can be described in the neighborhood
of the $a$-th NS5 brane in terms of a set of anti-fundamental chirals coupled to the $(a-1)$-th gauge group, and
fundamental chirals coupled to the $a$-th gauge group. Integrating out the chiral multiplets, and
evaluating the displacement operator we obtain
\begin{equation}
r_{(a)}^+(m) = \tilde R_a(m) - \tilde R_{a-1}(m)
\end{equation}
in terms of the vevs of the $\CN=1$ chiral operators for the $(a-1)$-th and $a$-th gauge groups.

We put a subscript in $r_{(a)}^+(m)$ because we cannot exclude the possibility that the different dual descriptions of the same
surface defect may be related up to some overall shifts, such as the $\frac{1}{2} a_D + m \log \Lambda^2$
twisted effective superpotential shift we encountered between the two descriptions of the simple surface defect in pure $SU(2)$.
Such shifts in $\CW$ may translate into shifts of the displacement operators.

In order to match the brane result, we propose that the unified relation
\begin{equation} \label{eq:wr}
v = m \qquad w(v) = r^+(m) = f_a(m)+ \tilde R_a(m) - \tilde R_{a-1}(m) ~,
\end{equation}
which includes the classical effect of the deformation in the NS5 brane and the effect of the D4 branes back-reaction.
Thus we learn how to relate the brane solution
 and the expectation values of the $\CN=1$ chiral operators for the $\CN=1$
deformed linear quiver gauge theory.

We expect it would be straightforward to use this identification to make new predictions for the linear or circular quiver gauge theories.
It is clear, for example, that the $\tilde R_1(m)$ expectation value for the leftmost gauge group in a linear quiver with $L-1$ nodes
will be the value on the first sheet of an $L$-sheeted function on the $m$ plane, whose values on the other sheets
will be identified with certain differences of expectation values of the chiral operators for consecutive gauge groups.
We also expect that these answers could be reproduced by a careful analysis of anomaly equations, or of
appropriate matrix models \cite{Dijkgraaf:2002dh}, using similar strategies as in the analysis of the $\CN=2$ matrix models in \cite{Nekrasov:2012xe}.

\subsection{Displacement operators in the $(2,0)$ six-dimensional SCFT} \label{sec:twozero}
For completeness, it is useful to look at the higher dimensional origin of displacement operators
for surface defects in general class $\CS$ theories. Half-BPS surface (i.e.\  codimension four) defects
in the six-dimensional $(2,0)$ SCFTs are analogue of Wilson line defects in maximally supersymmetric gauge theories,
and reduce to them upon compactification on a circle.

A supersymmetric Wilson loop in 5d SYM
\begin{equation}
P\exp \int_{- \infty}^{\infty} (A + i \phi^5)
\end{equation}
breaks the $SO(5)_R$ symmetry group to an $SO(4)_R$ subgroup.
Acting on the Wilson line defect with broken R-symmetry generators
gives an insertion of one of the other four scalar fields
\begin{equation}
\left[ P\exp \int_{- \infty}^{0} (A + i \phi^5) \right] \phi^A \left[ P\exp \int_{0}^{\infty} (A + i \phi^5) \right]
\end{equation}
along the Wilson line. These combine with other similar operators, e.g.\  the insertion of certain fermions or of field strengths, to form the displacement multiplet.
Notice that this displacement operator satisfies interesting chiral ring relations: if we take some complex combination such as $\phi^1 + i \phi^2$
and raise it to a sufficiently high power, we can rewrite it in terms of smaller powers multiplied by traces $\Tr (\phi^1 + i \phi^2)^k$, which can be taken out of the line defect and identified with
the value of Coulomb branch bulk operators brought to the defect.

In a similar manner, the surface defects in six dimensions also break the $SO(5)_R$ symmetry group to an $SO(4)_R$ subgroup, and
thus we have an $SO(4)_R$ vector of displacement operators, in an appropriate supermultiplet of the unbroken
$OSp(4^*|2) \times OSp(4^*|2)$ superconformal symmetry. We expect them to satisfy appropriate chiral ring relations involving the
vevs of Coulomb branch bulk operators. In particular, at a locus of Coulomb branch where the theory reduces to an Abelian theory
on a multi-sheeted cover of space-time, the displacement operators vevs should coincide with the vevs which describe the motion of the M5 brane
in the directions  transverse to the defect.

Upon twisted compactification on a Riemann surface, the six-dimensional theory flows to a four-dimensional class $\CS$ theory. A surface defect at a point on the Riemann surface
flows to a surface defect in that theory. Two of the six-dimensional displacement operators, $(2,2)$ superpartners of a displacement along the Riemann surface,
have a vev that should coincide with the motion of the M5 branes along the cotangent bundle to the Riemann surface. This is the twisted chiral operator we denoted as $\sigma$ in the field theory examples.
The other two six-dimensional displacement operators, $(2,2)$ superpartners of a displacement in space-time, have a vev that should coincide with the motion of the M5 brane in the $w$ direction,
as we saw in the brane examples.

\section{Conclusions and Outlook}
The methods described in this note allow one to derive the low energy effective twisted superpotential for any surface defect defined by coupling a $2d$ GLSM
to a four-dimensional gauge theory, provided that one is equipped with the resolvents appropriate for the representation content of the $2d$ GLSM.
The resolvents for the fundamental representation are available for all linear quiver gauge theories with unitary gauge groups
thanks to the recent localization calculations of \cite{Nekrasov:2012xe}.
It would be interesting to derive the resolvents for other representations
and gauge groups, through localization, anomaly equations or brane engineering.

Our calculations suggest a novel interpretation of the classical integrable systems that appear in $\CN=2$ theories.  Their coordinates are the parameters of general GW surface defects in the theory. We derived this interpretation for the periodic Toda system associated to the pure $SU(N)$ gauge theory. It would be interesting to repeat the exercise for
other examples such as elliptic Calogero-Moser. It should also be possible to repeat our calculations for codimension two defects in five-dimensional gauge theories compactified on a circle.

Finally, we have also initiated a study of how $\CN=1$ deformation of the bulk theory affect the surface defects. We have seen strong evidence for non-renormalization theorems
protecting some aspects of the surface defect dynamics and used the surface defects to establish a link between a brane analysis of the problem and the anomaly equations analysis.

An interesting puzzle raised by this paper is the physical, field-theoretic origin of the hidden vacua which are reached by analytic continuation through the cuts in the resolvents.
The nature of the hidden vacua appears to be rather universal, only weakly affected by the matter content of the bulk gauge theory.
Some of these features appear to survive $\CN=1$ deformations of the bulk theory.

\section*{Acknowledgements}
We are grateful to A.~Kapustin for collaboration at an early stage of the project, and many important discussions.
The research of DG was supported by the Perimeter Institute for Theoretical Physics.
Research at Perimeter Institute is supported by the Government of Canada through Industry Canada and by the Province of Ontario through the Ministry of Economic Development and Innovation.
The work of SG is supported in part by DOE Grant DE-FG03-92-ER40701FG-02.
The work of NS was supported in part by DOE grant DE-SC0009988 and by the United States-Israel Binational Science Foundation (BSF) under grant number~2010/629.
Opinions and conclusions expressed here are those of the authors and do not
necessarily reßect the views of funding agencies.

\appendix
\section{Manipulations of the flag manifold defect chiral ring equations} \label{app:man}
Without loss of generality, we can write the constraint as
\begin{equation}
P_N(z) - e^{t_{K-1}} Q_{K-2}(z) - \frac{\Lambda^{2N} e^{-t_{K-1}}}{Q_{K-2}(z)} = \frac{Q_{K-1}(z) \tilde Q(z)}{Q_{K-2}(z)} ~,
\end{equation}
but this relation does not look particularly instructive.

We can improve on this expression if we borrow a bit more material from the theory of continued fractions.
We can consider a second solution $X_k(z)$ of the recursion relation
\begin{equation}
X_{k+1}(z) - e^{t_k} X_{k-1}(z) = X_k(z) q_k(z)
\end{equation}
and observe that
\begin{equation}
X_{k+1} Q_k - Q_{k+1} X_k =  e^{t_k} \left(X_{k-1} Q_k - Q_{k-1} X_k\right)
~. \end{equation}
Thus if we start the recursion from $X_0(z) = 0$ and $X_1(z) = 1$, then
\begin{equation}
X_{K-1} Q_{K-2} - Q_{K-1} X_{K-2} = (- e^{t_{K-2}})(- e^{t_{K-3}})\cdots (- e^{t_{2}})(- e^{t_{1}})
~. \end{equation}
Thus if we can solve the rational part of the constraint by setting
\begin{equation}
\tilde Q(z) = Q_{K-2}(z) q_{K-1}(z)- \Lambda^{2N} X_{K-2}(z) \prod_k (- e^{-t_{k}}) ~,
\end{equation}
so that
\begin{equation}
\frac{Q_{K-1}(z) \tilde Q(z)}{Q_{K-2}(z)} = Q_{K-1}(z) q_{K-1}(z) - \Lambda^{2N} X_{K-1}(z) \prod_k (- e^{-t_{k}})- \frac{\Lambda^{2N} e^{-t_{K-1}}}{Q_{K-2}(z)}~. \end{equation}
Thus we arrive at a neater form of the relation
\begin{equation}
P_N(z) - e^{t_{K-1}} Q_{K-2}(z) = Q_{K-1}(z) q_{K-1}(z) - \Lambda^{2N} X_{K-1}(z) \prod_k (- e^{-t_{k}})
~. \end{equation}

This equation differs from the ungauged case only by the extra term proportional to $X_{K-1}(z)$. We can also write
$X_{K-1}(z)$ as the determinant of a matrix $\tilde A(z)$, which is obtained from $A(z)$ by removing the first and last
rows and columns, so that
\begin{equation}
P_N(z)= \det A(z) - \Lambda^{2N} \tilde A(z) \prod_k (- e^{-t_{k}})
~. \end{equation}
A final manipulation allows us to reabsorb the extra term by a modification of the $A(z)$ matrix:
\begin{equation}
\hat A(z,t) = \begin{pmatrix} q_0(z) & -1 & 0 & \cdots & 0 & 0 & -\Lambda^{2N} e^{-t}\prod_k (- e^{-t_{k}}) \cr
e^{t_1} & q_1(z) & -1& \cdots & 0 & 0 & 0 \cr
\cdots & \cdots & \cdots & \cdots & \cdots & \cdots & \cdots \cr
0 & 0 & 0 &\cdots & e^{t_{K-2}} & q_{K-2}(z) & -1 \cr
- e^t & 0 & 0 & \cdots & 0 & e^{t_{K-1}} & q_{K-1}(z) \end{pmatrix}
\end{equation}
so that
\begin{equation}
\det \hat A(z,t) = P_N(z) -e^t - \Lambda^{2N} e^{-t}
~. \end{equation}

\bibliographystyle{JHEP_TD}
\bibliography{surface}

\end{document}